\documentclass[aps,prb,amsmath,amssymb,twocolumn,superscriptaddress,floatfix,longbibliography]{revtex4-2}

\usepackage{dcolumn}%
\usepackage{bm}%

\usepackage[justification=justified]{subcaption}
\usepackage{graphicx}

\usepackage[utf8]{inputenc}
\usepackage[T1]{fontenc}
\usepackage{braket}
\usepackage{color}

\usepackage{csquotes}

\newcommand{\nid}{\noindent}
\usepackage{acronym}
\acrodef{RPMD}[RPMD]{Ring Polymer Molecular Dynamics}
\acrodef{NAE}[NAE]{Non adiabatic effects}
\acrodef{NAE}[NAE]{Non adiabatic effects}
\acrodef{SI}[SI]{Supporting Information}
\usepackage[utf8]{inputenc}
\usepackage[T1]{fontenc}
\usepackage{hyperref}
\hypersetup{
colorlinks=true,
urlcolor= blue,
citecolor=blue,
linkcolor= blue,
}
\usepackage{tabularx}
\usepackage{booktabs}
\usepackage{orcidlink}
\usepackage{etoolbox}
\usepackage{lipsum}  
\usepackage{soul}
\usepackage[font=small]{caption}
\newcommand{\rev}[1]{{\color{black}#1}}

\usepackage{caption}

\makeatletter
\renewcommand\@makecaption[2]{%
  \par
  \vskip\abovecaptionskip
  \begingroup
   \small\rmfamily
    \begingroup
     \samepage
     \flushing
     \let\footnote\@footnotemark@gobble
     \@make@capt@title{#1}{#2}\par
    \endgroup
  \endgroup
  \vskip\belowcaptionskip
}
\makeatother

\begin{document}

\preprint{AIP/123-QED}

\title{Quantum rates in dissipative systems with spatially varying friction
}

\author{Oliver Bridge \orcidlink{0009-0008-6835-5082}}%
\affiliation{Yusuf Hamied Department of Chemistry,  University of Cambridge,  Lensfield Road,  Cambridge,  CB2 1EW, UK}

\author{Paolo Lazzaroni}%
\affiliation{MPI for the Structure and Dynamics of Matter, Luruper Chaussee 149, 22761 Hamburg, Germany \looseness=-1}

\author{Rocco Martinazzo \orcidlink{0000-0002-1077-251X}}%
\affiliation{Department of Chemistry, Università degli Studi di Milano, Via Golgi 19, 20133 Milano, Italy}

\author{Mariana Rossi \orcidlink{0000-0002-3552-0677}}%
\affiliation{MPI for the Structure and Dynamics of Matter, Luruper Chaussee 149, 22761 Hamburg, Germany \looseness=-1}

\author{Stuart C. Althorpe}%
\affiliation{Yusuf Hamied Department of Chemistry,  University of Cambridge,  Lensfield Road,  Cambridge,  CB2 1EW, UK}

\author{Yair Litman \orcidlink{0000-0002-6890-4052}}%
\email[Email: ]{yl899@cam.ac.uk}
\affiliation{Yusuf Hamied Department of Chemistry,  University of Cambridge,  Lensfield Road,  Cambridge,  CB2 1EW, UK}

\begin{abstract}

We investigate whether making the friction spatially dependent \rev{on the reaction coordinate} introduces quantum effects into the thermal reaction rates for dissipative reactions. \rev{Quantum rates are calculated}  using the numerically exact multi-configuration time-dependent Hartree (MCTDH) method, as well as the approximate ring-polymer molecular dynamics (RPMD), ring-polymer instanton (RPI) methods, and classical molecular dynamics. 
By conducting simulations across a wide range of temperatures and friction strengths, we can identify the various regimes that govern the reactive dynamics. At high temperatures, in addition to the {spatial-diffusion} and  {energy-diffusion} regimes predicted by Kramer's rate theory, a (coherent) {tunnelling-dominated} regime is identified at low friction. At low temperatures, incoherent tunnelling dominates most of Kramer's curve, except at very low friction when coherent tunnelling becomes dominant.
Unlike in classical mechanics, the bath's influence changes the equilibrium time-independent properties of the system, leading
to a complex interplay between spatially dependent friction and nuclear quantum effects even at high temperatures. %
More specifically, a realistic friction profile can lead to an increase (decrease) of the quantum (classical) rates with friction within the {spatial-diffusion} regime, showing that classical and quantum rates display qualitatively different behaviours.
Except at very low frictions, we find that RPMD captures most of the quantum effects in the thermal reaction rates.

\end{abstract}

\maketitle

\section{\label{sec:Introduction} Introduction}

The accurate modelling of chemical rates in dissipative environments is of fundamental importance to chemistry, physics, and biology \cite{Kramers_1940, Hanggi_RevModPhys_1990,Pollak_CPC_2023}. In many systems, the reactive process can be approximated by the time evolution of a reaction coordinate coupled to a thermalizing and fluctuating environment.  A simple realization of this picture is obtained with system-bath models, where the system coordinates are %
coupled to numerous harmonic bath degrees of freedom and the average dynamics of the system is fully governed by the bath temperature and the friction kernel \cite{Weiss_book,Tuckerman}. The first step towards accurate theoretical predictions is thus to ensure that the friction kernel is an appropriate representation of the underlying dynamics.

A key assumption often made in constructing system-bath models is to neglect the influence of the reaction coordinate  on the friction kernel by linearizing the system-bath coupling. 
However, this approximation proves inaccurate for a variety of dynamical processes, including simple Lennard-Jones fluids
\cite{Cukier_JCP_2008, Posch_MolPhys_1981}, proton-transfer reactions in the condensed phase
\cite{Antoniou_JCP_1999}, and adsorption at interfaces 
\cite{Polley_2024}. Another interesting instance of position-dependent friction occurs in the dynamics of atoms and small molecules near metals.  In these systems, the movement of the nuclei can induce non-equilibrium fluctuations of the electrons within the metal, generating (electronic) frictional forces \cite{Dou_JCP_2018, Rocco_PRA_2022, Rocco_PRL_2022,Martinazzo_2023} that significantly modify the nuclear dynamics \cite{Dou_JCP_2016, Spiering_JPCL_2019}. This type of spatially-dependent friction (SDF) is especially relevant when molecules approach interfaces, since the frictional force goes from zero in the vacuum to a finite value at the metal \cite{Maurer_PRL_2017, Spiering_JPCL_2019}.

Several studies have highlighted significant deviations between linear and non-linear models \cite{Carmeli_CPL_1983, Straus_JCP_1993, Brown_JCP_1995,Voth_JCP_1992,Krishnan_JCP_1992,Haynes_CPL_1993, Haynes_JCP_1994}.
For example, Pollak-Grabert-H\"anggi (PGH) theory has been extended to address
SDF \cite{Krishnan_JCP_1992,Haynes_CPL_1993,Haynes_JCP_1994}, and Voth has developed an 
effective Grote-Hynes theory \cite{Voth_JCP_1992}.
In the strong-damping limit, both approaches calculate an average spatial modification of the friction coefficient near the barrier top and introduce it in expressions derived from position-independent theories. In this way, it has been found that a spatial  antisymmetric reaction-coordinate dependence of the friction profile
around the transition state leads to a negligible modification of chemical reaction rates  \cite{Haynes_JCP_1995}, even when the coupling is strongly non-linear.

In most of these studies, nuclei are considered to be classical (Newtonian) particles. However, tunnelling and zero-point energy can change the reaction rate by several orders of magnitude and fundamentally modify its temperature dependence. It is generally accepted that coupling to a bath
diminishes the magnitude of nuclear quantum effects (NQEs) \cite{Wolynes_PRL_1981, Weiss_book}. \rev{
Most of the previous work on system-bath models, including formal analytical approaches \cite{Hanggi_RevModPhys_1990,Pollak_CPC_2023} and numerically accurate quantum calculations \cite{Makri_JCP_1998, Wang_JCP_2006,Craig_JCP_2007}, have been limited to position-independent friction. As a result, the effects of SDF on quantum reaction rates remained overlooked with a few important exceptions. 
At high temperatures, approximate numerical studies  reported by Navrotskaya and Geva have shown
that a non-bilinear system-bath coupling can lead to an enhancement of quantum reaction rates \cite{Navrotskaya_ChemPhys_2006}, and 
Antoniou and Schwartz reported 
a reduction of kinetic isotope effects in similar scenarios \cite{Antoniou_JCP_1999}. At low temperatures,  some of the present authors reported recently that an SDF profile can steer nuclear tunnelling at low temperatures \cite{Litman_JCP_2022b}. 
Thus, a systematic study of the interplay of SDF and quantum dynamics over extended friction and temperature regimes has remained elusive.
}

 In this article, we calculate quantum reaction rates for representative system-bath models with SDF, in regimes ranging from activated \rev{barrier crossing} to deep tunnelling. We report numerically exact quantum rates at finite temperatures and use approximate methods based on imaginary-time path integrals to rationalize the results and elucidate the different rate-determining mechanisms for quantum and classical rates.

\section{\label{sec:Methods} Theory and Methods}

\subsection{Calculation of Thermal Rate Constants}

Let us consider a system that can be divided into a reactant and a
product region. The quantum thermal rate constant, $k_\text{Q}(T)$, of a generic "reactant"$\to$"{product}" reaction can be expressed as \cite{Yamamoto1960,Miller1983,Miller1998} %
\begin{equation}
\begin{split}
   k_\text{Q}(T)  =
      \frac{1}{Q_r(\beta)} c_\text{fs}(\beta,t) \bigg\vert_{t>\tau},
\end{split}\label{eq:rate1}
\end{equation}
\nid where $\beta = (k_BT)^{-1} $ is the inverse temperature, $Q_r(\beta)=\text{Tr}[e^{-\beta \hat{H}}(1-\hat{h})]$ is the reactant partition function, $\hat{H}$
is the Hamiltonian of the system,  and
$\hat{h}$ is the quantum projection operator onto the product states. The flux-side correlation function is defined as 
\begin{equation}
\begin{split}
   c_\text{fs}(\beta,t) = \text{Tr}[e^{-\beta \hat{H}/2} \hat{F} e^{-\beta \hat{H}/2} \hat{h}(t) ],
\end{split}\label{eq:rate2}
\end{equation}
\nid where $\hat{F}=\frac{i}{\hbar}[\hat{H},\hat{h}]$ is the flux operator and $\hat{h}(t)=\exp(i\hat Ht/\hbar)\hat h\exp(-i\hat Ht/\hbar)$ at time $t$.

The $t>\tau$ condition in Eq.~\ref{eq:rate1} should be interpreted as the plateau time by which the correlation function reaches a constant value.  The existence of such a plateau requires a well-defined separation of timescales such that thermal fluctuations that take particles out of the wells to the barrier top are rare events in comparison with the rapid short-time dynamics that relax $c_\text{fs}(\beta,t)$ to the plateau \cite{Chandler_JCP_1978}.  
A refined formula for cases when the free-energy barrier is comparable to $k_B T$ was derived in Ref.~\onlinecite{Drozdov2001,Craig_JCP_2007}.

\subsection{\label{sec:Methods-MCTDH}Multi-configuration Time-dependent Hartree}

The multi-configuration time-dependent Hartree (MCTDH) method is a wavefunction method for solving the time-dependent Schrödinger equation for multidimensional systems composed of distinguishable particles \cite{mey90:73,bec00:1,mey09:book}. It employs a time-dependent basis to expand optimally (in a variational sense) the system wave function, $\Ket{\Psi(t)}$,  and thus mitigate the exponential scaling problem affecting standard basis set methods. Specifically, the MCTDH %
 wavefunction takes the form of a linear combination of  products of single-particle functions (SPFs) $\ket{\phi^{(k)}_{j}(t)}$, one for each "mode" $k$,
\begin{equation}
\begin{split}\label{eq:MCDTDH1}
\Ket{\Psi(t)} = \sum_{J}A_J(t) \prod_k \ket{\phi_{j_k}^{(k)}(t)},
\end{split}
\end{equation}
\nid and both the expansion coefficients  $A_J$ and the SPFs are time-evolved according to variational equations of motion \cite{mey90:73,bec00:1,mey09:book}. In the above expression 
$J=(j_1,...,j_k,...,j_F)$ is a multi-index and $j_k=1,...,n_k$ labels the SPFs used for the $k^{\textit{th}}$ mode. The modes represent either single degrees of freedom of the system or a group thereof. 
 In the multi-layer extension \cite{mey90:73,bec00:1,mey09:book} of MCTDH (known as ML-MCTDH) the SPFs are taken to be high-dimensional and are further expanded in MCTDH form employing lower-dimensional SPFs, which in turn can be similarly expanded. This generates a hierarchical construction, a "multi-layer tree", which is terminated with low-dimensional SPFs that are directly expanded on a grid or a basis-set (the so-called primitive grid).
This gives the wavefunction a rather flexible structure which makes possible the treatment of 
quantum systems with several hundred degrees of freedom, provided the Hamiltonian takes a relatively simple form \cite{Kondov_JPCC_2007,Westermann_JCP_2011}.

ML-MCTDH has been successfully applied to the calculation of quantum thermal rate constants
in condensed-phase problems \cite{ML-MCTDH_rates,Craig_JCP_2007,Litman_JCP_2022b}. In these calculations, the traces appearing in   $Q_r(\beta)$ and in Eq.~\ref{eq:rate2}, $c_\text{fs}$, are evaluated stochastically. %
This entails their replacement with an incoherent sum over a finite number (typically some hundreds) of representative elements from the Hilbert space of the system, which are selected stochastically using an importance sampling technique and later handled with ML-MCTDH.
Specifically, the resulting "Monte-Carlo wavepacket" procedure can be summarized as follows. In the first step, a sample of pure states $\{\ket{\Phi_i}\}_i$ is drawn from the thermal equilibrium state of the (uncoupled) bath, and is combined with special system states $\{\ket{\phi_\nu}\}_\nu$ to form representatives of the total system, $\ket{\Psi_I}=\ket{\Phi_i}\ket{\phi_\nu}$. In the second step, the $\ket{\Psi_I}$ are %
propagated in imaginary-time using the full Hamiltonian so that they thermalize at the given temperature, $\ket{\Psi_I^\beta}=e^{-\beta \hat H/2} \ket{\Psi_I}$. Finally, the $\ket{\Psi_I^\beta}$  are propagated in real-time with the same Hamiltonian, $\ket{\Psi_I^\beta}\rightarrow\ket{\Psi_I^\beta(t)}$, and used to compute appropriate expectation values from which the flux-side correlation function can be easily obtained. Calculation of the reactant partition function proceeds  similarly, and is simpler, since it does not require any real-time propagation. This is the strategy developed by Craig \emph{et al.} in Ref.~\onlinecite{Craig_JCP_2007}, where the interested reader can find the necessary details. Here, we use mainly the implementation described in our previous article, Ref.~\onlinecite{Litman_JCP_2022b}, with a few modifications that are described in Sec.~\ref{sec:Simulation_details-MCTDH}.

\subsection{Ring Polymer Molecular Dynamics}

Ring polymer molecular dynamics (RPMD) is an approximation rooted in the (imaginary-time) path integral formulation of quantum mechanics. It utilizes a classical time evolution in an extended ring-polymer phase space to approximate the effects of quantum thermal fluctuations on the dynamics of condensed-phase systems \cite{Craig_JCP_2004, Habershon_AnnRev_2013}.

In the following, we consider the ring polymer Hamiltonian of an $F+1$-dimensional system given by
\begin{equation}
\begin{split}\label{eq:H_P}
   H_P (\mathbf{p} , \mathbf{q}) &= U_P(\mathbf{q}) +
  \sum_{j=0}^{F} \sum_{k=1}^P  \frac {\left( p_j^{(k)}\right)^2 }{2m_j},
\end{split}
\end{equation}
where 
\begin{equation}
\begin{split}\label{eq:U_P}
  U_P(\mathbf{q}) =& \sum_{j=0}^{F} \sum_{k=1}^P \left[
   \frac{m_j\omega_P^2}{2} \left({q}_j^{(k)}-{q}_j^{(k+1)}\right)^2 \right]  + \\&  
   \sum_{k=1}^P V\left( q_0^{(k)},  q_1^{(k)}, \dots,  q_{F}^{(k)}\right).
\end{split}
\end{equation}
and  ${\bf q}^{(k)}=\{q^{(k)}_0,q^{(k)}_1,\dots,q^{(k)}_F\}$ 
represent the positions of the $k$-th ring-polymer bead, $\mathbf{q}=\{{\bf q}^{(1)},{\bf q}^{(2)},\dots, {\bf q}^{(P)}\}$ is a short form denoting all the position coordinates,  ${\bf p}^{(k)}$ and $\mathbf{p}$ are similarly defined for the momenta, $m_j$ is the mass of the $j$-th degree of freedom, $V$ is the potential energy surface (PES), and
$\omega_P = (\beta_P\hbar)^{-1}$, with $\beta_P = \beta/P$. 
The RPMD approximation to the exact quantum rate constant is given by \cite{Craig_JCP_2005, Craig_JCP_2005_2}
\begin{equation}
\begin{split}
   k_\text{Q}(T) \approx k_\text{RPMD}(T)&=\lim_{P\to \infty} k^{(P)}(T) \\ &= 
     \lim_{P \to \infty}  
      \frac{1}{Q_r^{(P)}} c_\text{fs}^{P}(\beta,t)\bigg\vert_{t>\tau},
\end{split}\label{eq:rate_RPMD}
\end{equation}
\nid where $\tau$ is the plateau time beyond which $c_\text{fs}^{P}(\beta,t)$ has become time-independent. Here, the  reactant partition function, $Q_r^{(P)}$, is the $P$-bead path-integral approximation to its exact quantum counterpart, and 
\begin{equation}
\begin{split}
   c_\text{fs}^{P}(t) = \frac{1}{(2\pi\hbar)^{F P}} \int\text{d}\bm{p}
   \int\text{d}\bm{q}\:\text{e}^{-\beta_P H_P} \delta[s(\bm{q})]v_\text{s}(q) h[s(\bm{q}(t)]
\end{split}
\end{equation}
is the ring-polymer flux-side time-correlation function, in which $s({\bf q})$ is a dividing surface between reactant and products,
$v_\text{s}$ is the velocity component orthogonal to $s$, and $h$ is a heaviside step function. The time evolution of ${\bf q}(t)$ is generated using the classical equations of motion obtained from the ring-polymer Hamiltonian of Eq.~\ref{eq:H_P}.

The RPMD rate constant, $k_\text{RPMD}(T)$, is clearly artificial but it has a number of important properties which ensure that it is often a good approximation to the exact quantum rate: i) it is exact in the harmonic and classical limits \cite{Craig_JCP_2005, Craig_JCP_2005_2}, ii) it is independent of the position of the dividing surface $s$, and iii) if $s$ is constructed to be invariant under cyclic permutation of the polymer beads, the corresponding RPMD TST rate, $k_\text{RPMD}^\ddag(T)$, (obtained by taking $\tau\to 0^+$ in Eq.~\ref{eq:rate_RPMD}) gives the exact quantum flux through the dividing surface in the limit $t\to 0^+\,$ \cite{Hele_JCP_2013}, which correctly accounts for the dominant effects of instantons in the deep-tunnelling regime \cite{JOR_JCP_2009}. If the barrier is symmetric (as is the case for the systems treated here \footnote{In the asymmetric friction model the system-bath is not symmetric about the barrier, so the optimal dividing surface could potentially mix in non-centroid ring-polymer modes at sufficiently low temperatures.}), the optimal dividing surface $s({\bf q})$ is a function of just the ring-polymer centroid $\mathbf{q}^\text{c}=\{q^{c}_0,q^{c}_1,\dots,q^{c}_F\}$, which has components \begin{equation}\label{eq:centroid}
\begin{split}
   q_{j}^\text{c} = \sum_{k=1}^P q_j^{(k)},
\end{split}
\end{equation}
and in this case $k_\text{RPMD}^\ddag(T)$ is identical to the centroid-TST rate \cite{Gillan_PRL_1987,voth1989}.
 Because of its computational efficiency, RPMD has been used to compute rates in complex systems \cite{Lawrence_FarDis_2020, Habershon_AnnRev_2013}, including polyatomic gas-phase  reactions \cite{Collepardo_JCP_2009, Suleimanov_JCP_2011, Huo_JPCL_2021,Allen_JCP_2013,Hickson_JPCL_2015, Suleimanov_JPCA_2016} and protein rearrangement reactions \cite{Boekelheide_PNAS_2011}.

\subsection{Ring Polymer Instanton Method}

The ring polymer instanton (RPI) method~\cite{JOR_JCP_2009, Arni_thesis} 
 is an  efficient semi-classical method for computing tunnelling rates in the "deep tunnelling" regime
\begin{equation}
      \label{eq:Tc}
 T<T_c^\circ = \frac{\hbar \omega^\ddag}{2\pi},
\end{equation}
in which the saddle point on $U_P({\bf q})$ is delocalised into an
 imaginary-time periodic orbit  known as the "instanton" \cite{Miller_JCP_1975,Jor_review_2018}. \rev{We note that this expression applies to barriers that resemble parabolas close to the barrier top; for flatter barriers, the determination of $T_c$ is not so simple~\cite{Fang_PRL_2017,McConnell_JComChem_2017}}. In the RPI method, the instanton is located by running  standard
saddle point search algorithms on $U_P({\bf q})$ \cite{Rommel_JCTC_2011,Nichols_JCPS_1990,Litman_thesis}. The large computational cost associated with the sampling procedure is thus replaced by a few geometry optimization and Hessian calculations. The tunnelling rate can be expressed as \cite{Jor_review_2018}
\begin{equation}
\begin{split}
   \label{eq:Kinst1}
   k_\text{inst}(T) = 
   \frac{1}{Q_r^{(P)}}
    \frac{1}{\beta_P \hbar}
\sqrt{ \frac{B_P(\mathbf{\bar{q}})}{2 \pi \beta_P \hbar^2}}
    Q_\text{vib} 
    e^{-S_P(\mathbf{\bar{q}})/\hbar} ,
       \end{split}
\end{equation}
\nid where %
\begin{equation}
      \label{eq:S_P}
   S_P/\hbar = \beta_P U_P ,
\end{equation}
$\mathbf{\bar{q}}$ is the position vector corresponding to the optimized instanton geometry, which
\nid can be identified as the discretized version of the Euclidean action with $U_P$ given by Eq.~
\ref{eq:U_P}, and
\begin{equation}
\begin{split}
   \label{eq:BP}
   B_P ({\bar{q}})= \sum_{i=1}^{F} \sum_{k=1}^P m_i({\bar{q}}^{(k+1)}_i-{\bar{q}}^{(k)}_i)^2.
\end{split}
\end{equation}
\nid The instanton vibrational partition function, $Q_\text{vib}$, is approximated  by 
\begin{equation}
\begin{split}
   \label{eq:Qvib}
Q_\text{vib} = \prod_k  \frac{1}{\beta_P\hbar|\lambda_k|},
\end{split}
\end{equation}
\nid where $\lambda_k$ are the non-zero eigenvalues of the ring-polymer dynamical matrix \cite{Beyer_JPCL_2016}.

 The RPI method has been successfully applied to systems containing hundreds of atoms using accurate  \textit{ab initio} potential energy surfaces  \cite{Rommel_JPCB_2012,Beyer_JPCL_2016,Litman_JACS,Litman_PRL, Fang_NatCom_2020}. The RPI rate $k_\text{inst}(T)$ typically agrees with the exact quantum rate to within a factor of two \cite{Jor_review_2018}.%

\section{Simulation details\label{sec:Simulation_details}}

\subsection{ System-Bath Model with Position-dependent Friction}

We consider a system-bath model described by the following PES \cite{Topaler_JCP_1994,Craig_JCP_2005}
\begin{equation}
\begin{split}
V(q;x_1,\dots,x_F) &= V_\text{sys}(q)+
\sum_{j=1}^{F}   \frac{m_j\omega_j^2}{2} \left(x_j - \frac{c_j g(q)}{ m_j \omega_j^2}\right)^2,
\end{split}\label{eq:Vsysbath}
\end{equation}
\nid where $q$ and $\{x_1,\dots,x_F\}$ correspond to the (1D) system and bath degrees of freedom, respectively, and $V_\text{sys}$ refers to the PES of the system in the absence of a bath. The bath is described by a %
spectral density 
\begin{equation}
\begin{split}\label{eq:J}
J(q,\omega) =
\left(\frac{\partial g(q)}{\partial q}\right)^2
\frac{\pi}{2}
\sum_{j=1}^{F} 
\frac{c_j^2}{m_j\omega} (\delta(\omega-\omega_j)+\delta(\omega+\omega_j)),
\end{split}
\end{equation}
\nid where $g(q)$ determines the position-dependence of the system-bath coupling. %
Equivalently, the system-bath coupling can be %
subsumed in a position- and frequency-dependent friction kernel, $\tilde{\eta}(q, \lambda)$, which is related to the bath spectral density by 
\begin{equation}
\begin{split}\label{eq:eta_laplace}
\tilde{\eta}(q, \lambda) = \frac{1}{\pi} \int_{-\infty}^{\infty}\text{d}\omega \frac{J(q,\omega)}{\omega}\frac{\lambda}{\omega^2+\lambda^2},
\end{split}
\end{equation}
and is position-independent when $g(q)=q$.
\begin{figure}%
    \includegraphics[width=0.95\columnwidth]{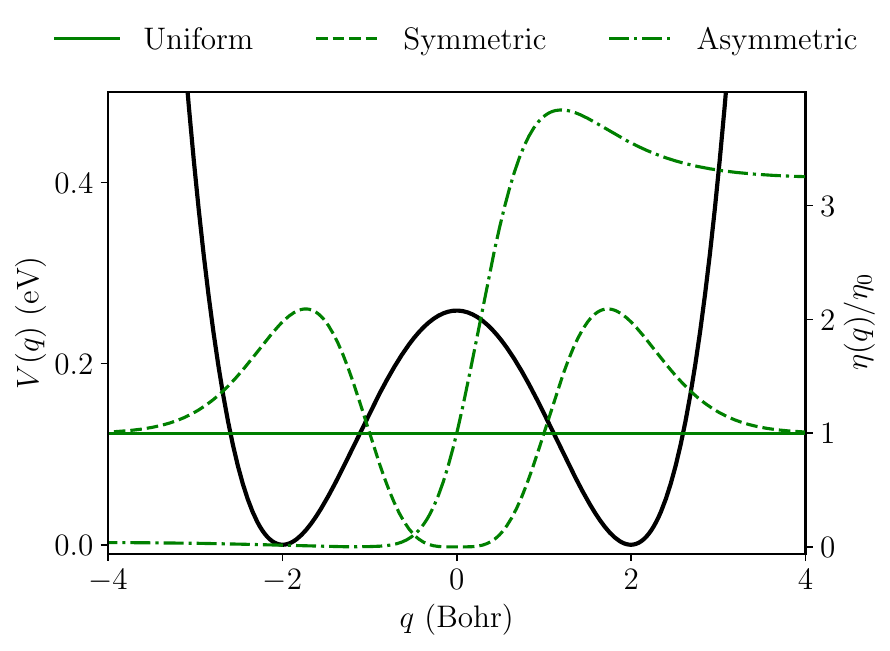}
    \caption{\label{fig:pes} Friction profiles for the uniform (solid green line), symmetric (dashed) and asymmetric (dot-dashed) models of Tab.~I,  with the DW1 potential energy (black line).}
\end{figure}
In this work, we use simple models for $V_\text{sys}$ and $\tilde{\eta}$.
For $V_\text{sys}$, we use the "DW1" double-well model of Topaler and Makri \cite{Topaler_JCP_1994},  
\begin{equation}
\begin{split}
V_\text{sys}(q) = - \frac{1}{2} m \omega^{\ddag 2} q^2 + \frac{m^2\omega^{\ddag 4}}{16V_0}q^4,
\end{split}\label{eq:DW}
\end{equation}
\nid where $\omega^\ddag=500$ cm$^{-1}$, $V_0=2085$ cm$^{-1}$, and  $m$ is the mass of atomic hydrogen.
For the  system-bath coupling, we use%
\begin{equation}
\begin{split}\label{eq:coupling_g}
g(q) = q [1 + \epsilon_1 \exp(-(q/\delta)^2/2) + \epsilon_2 \tanh(q/\delta) ],
\end{split}
\end{equation}
where $\delta$, $\epsilon_1$ and $\epsilon_2$ are positive real numbers. %
The spectral density is taken to be Ohmic with an exponential cutoff, 
\begin{equation}
\begin{split}
J^\text{Ohm}(q,\omega) = \bigg(\frac{\partial g(q)}{\partial q}\bigg)^2 \eta_0 \omega e^{-\omega/\omega_c},
\end{split}
\end{equation}
where $\eta_0$ is the Ohmic friction coefficient and $\omega_c=500$ cm$^{-1}$ is a frequency cutoff; when $g(q)=q$, $J^\text{Ohm}(q,\omega)$  reduces to the position-independent spectral density employed in the DW1 model of Ref.~\onlinecite{Topaler_JCP_1994}.

The factorization of the spectral density into position-dependent and frequency-dependent factors, usually referred to as separable coupling or uniform coupling \cite{Caldeira_AnPhys_1983, Haynes_CPL_1993},  assumes that the dynamical time scale is independent of the value of the reaction coordinate. 
This approximation is justified in most cases since, to lowest order, the system-bath and bath-bath couplings determine respectively the magnitude and the timescale of the friction kernel \cite{Haynes_CPL_1993}; it has also been shown numerically to give a good approximation to electronic friction in several metals \cite{Litman_JCP_2022b}.
\begin{table}[b]
    \centering
    \begin{tabular}{>{\centering\arraybackslash}p{2.2cm}|>{\centering\arraybackslash}p{1cm}>{\centering\arraybackslash}p{1cm}>{\centering\arraybackslash}p{1cm}}
        \toprule
         Model & $\epsilon_1$ & $\epsilon_2$& $\delta$\\
        \midrule
         Uniform       &    0.0    & 0.0 & 1.0   \\ 
         Symmetric \  &  -1.0     & 0.0 & 1.0    \\
         Asymmetric     &    0.0    & 0.8 & 0.5   \\
         \bottomrule
    \end{tabular}
    \caption{Parameters of the friction models employed in this work. }
    \label{tab:param}
\end{table}

We consider three friction models, one model with position-independent coupling, and two models with position-dependent coupling. The model parameters are summarized in  Tab.~\ref{tab:param}. 
In Fig.~\ref{fig:pes}, we plot the PES and friction coefficients along with the reaction coordinate. The uniform model has a constant friction profile and serves as a reference from which we can evaluate the impact that non-linear couplings (non-uniform friction) have on the thermal rate. The symmetric model  has a lower friction coefficient in the vicinity of the transition state and a larger one at the reactant and product wells. This friction profile resembles one recently constructed for hydrogen diffusion reactions in metals \cite{Litman_JCP_2022b}, for a proton-transfer reaction in liquid methyl chloride \cite{Antoniou_JCP_1999}, \rev{and the one studied by
Navrotskaya and Geva \cite{ Navrotskaya_ChemPhys_2006}}.
The asymmetric model has vanishing friction in the reactant well and larger friction in the product well, similar to the models employed by Straus and others \cite{Straus_JCP_1993,Haynes_CPL_1993,Navrotskaya_ChemPhys_2006}. \rev{Note that this model presents the same friction value as the uniform case at the transition state}.

\subsection{ML-MCTDH Calculations\label{sec:Simulation_details-MCTDH}}
The ML-MCTDH calculations were implemented using the Heidelberg package \cite{mctdh:package}. Except for the details given below, these calculations followed Ref.~\onlinecite{Litman_JCP_2022b} (Supplementary Information), which is an implementation of the Monte Carlo wavepacket strategy of Craig \emph{et al.} \cite{Craig_JCP_2007}.
The bath was described with $F = 50$  modes, using the logarithmic discretization of Ref.~\onlinecite{Wang_JCP_2001}, and its (uncoupled) thermal equilibrium state was sampled to extract between 256 and 512 realizations, $n_B$, for each temperature, $T$. These bath states were combined with $n_S$ system states to define a running sample of $n_B\times n_S$ wavepackets for each value of the temperature and of the coupling strength. The resulting $F+1$-dimensional wavepackets were propagated with ML-MCTDH, in both imaginary and real-time, using a single ML-tree structure. The latter was obtained after extensive tests at the extremes of the temperature--coupling strength intervals of interest (see Ref.~\onlinecite{Litman_JCP_2022b} [Supplementary Information] for details). The choice of the "bare" system states depends on the type of calculation---whether it is for the reactant partition function or the flux-side correlation function---and must be optimized to reduce the overall computational cost. This choice is particularly crucial for the flux-side calculation, since the flux operator $\hat F$ is intrinsically of high-rank, and could require many states if represented in the usual spectral form, $\hat F=\sum_\nu \nu \ket{\nu}\bra{\nu}$. The trick \cite{Craig_JCP_2007} is to observe that its Boltzmannized version $\hat F_\beta$ is of low rank, and therefore $\hat F$ is better rewritten as $\hat F=e^{\beta\hat  H_S/2}\hat F_\beta e^{\beta \hat H_S/2}$  where $\hat H_S$ is the system Hamiltonian. Typically, $\hat F_\beta$ is well represented by a small number of (time-reversed conjugate) pairs of eigenstates; in our calculations, $n_S=2$ was found to be sufficient, so that $\hat F_\beta\approx \nu_\beta \ket{\nu_\beta}\bra{\nu_\beta}-\nu_\beta \ket{\bar{\nu}_\beta}\bra{\bar{\nu}_\beta}$, where $\nu_\beta$ is the largest-magnitude eigen-flux and the bar denotes application of time-reversal. The drawback with this approach is that the reverse imaginary-time propagation needed to define the system states, $\ket{\phi_\beta}=e^{+\beta \hat H_S/2}\ket{\nu_\beta}$, can be numerically unstable due to high energy states in $\hat H_S$ which are, however, irrelevant for the dynamics.  To circumvent this problem, we exploit $\hat F e^{-\beta \hat H_S/2} =e^{\beta \hat H_S/2}\hat F_\beta$ to write $\nu_\beta \ket{\phi_\beta}=\hat F e^{-\beta \hat H_S/2}\ket{\nu_\beta}$, which only requires the numerically stable imaginary-time propagation $e^{-\beta \hat H_S/2}$ and the application of the bare flux operator $\hat F$. This amounts to the following formal manipulation of the flux operator  
\begin{align}
\hat F & =\hat Fe^{-\beta \hat H_{S}/2}\hat F_{\beta}^{-1}e^{-\beta \hat H_{S}/2}\hat F\\ \nonumber
 & \approx\sum_{\nu}\nu_{\beta}^{-1}\hat Fe^{-\beta \hat H_{S}/2}\ket{\nu_{\beta}}\bra{\nu_{\beta}}e^{-\beta \hat H_{S}/2}\hat F
\end{align}
where the sum is restricted to the largest-magnitude eigenvalues of the Boltzmannized flux operator; i.e., $\hat F_\beta^{-1}$ is replaced by the pseudo-inverse of a low-rank approximation of $\hat F_\beta$. 

\subsection{RPMD, Classical and RPI Simulations}

\rev{The ring-polymer potential used in RPMD and RPI simulations was obtained by substituting Eq. \ref{eq:Vsysbath} into Eq. \ref{eq:U_P}, such that each imaginary-time slice contains one system coordinate and the corresponding modes of the discretized bath}. The RPMD rate constants were computed using the Bennett-Chandler approach \cite{Chandler_JCP_1981,Bennett_book} which is based on the factorization
\begin{equation}
\begin{split}\label{eq:BC_rate}
    k_\text{RPMD}(T)=k_\text{RPMD}^\ddag(T;s)\kappa_\text{RPMD}(T;s),
\end{split}
\end{equation}
where  $\kappa_\text{RPMD}(T;s)$ is the RPMD transmission coefficient \cite{Collepardo_JCP_2009}  (i.e.\ the fraction of trajectories initiated at the dividing surface which remain on the product side at $t>\tau$). 
Note that $k_\text{RPMD}^\ddag(T;s)$ and $\kappa_\text{RPMD}(T;s)$ depend on the specific choice of the dividing surface, $s$,  whereas $k_\text{RPMD}(T)$ does not.
Unless specified otherwise, we take $s(\bm{q^\text{c}})=q_{0}^{c}-q_0^\ddag$, where $q_0^\ddag$ is the classical transition state and $q_{0}^{c}$ is the projection of the centroid along the system coordinate, so that $v_\text{s}=p_0^c/m_0$, where $p_0^c$ is the momentum conjugate to $q_{0}^{c}$. The terms on the right-hand side of Eq.~\ref{eq:BC_rate} can then be written
\begin{equation}
\begin{split}\label{eq:kp_TST_1}
k^\ddag_\text{RPMD}(T;s) = \bigg(\frac{k_B T}{2\pi m} \bigg)^{1/2}
\frac{\langle \delta(q_{0}^{c}-q_0^\ddag)\rangle}{\langle h(q_0^\ddag-q_{0}^{c})\rangle},
\end{split}
\end{equation}
\nid and
\begin{equation}
\begin{split}\label{eq:kp_RPMDTST}
\kappa_\text{RPMD}(T;s)=\kappa_\text{RPMD}(T,t;s)\bigg\vert_{t>\tau}
\end{split}
\end{equation}
\nid with
\begin{equation}
\begin{split}\label{eq:kp_RPMDTST_of_t}
\kappa_\text{RPMD}(T, t;s)=  
\frac{\langle \delta(q_{0}^{c}-q_0^\ddag)(p_0^c/m)h(q_{0}^{c}(t) - q_0^\ddag) \rangle}{\langle \delta(q_{0}^{c}-q_0^\ddag)(p_0^c/m) h(p_0^c)\rangle},
\end{split}
\end{equation}
where $\langle \dots \rangle$ denotes the average over the canonical ensemble determined by $H_p$ (see Eq.~\ref{eq:H_P}), and $\tau$ is the plateau time previously mentioned.

We computed $k^\ddag_\text{RPMD}$ by thermodynamic integration \cite{Collepardo_JCP_2009}, and $\kappa_\text{RPMD}$ by sampling from a thermal distribution with the ring-polymer centroid pinned at the barrier top. 
The spectral density was discretized using the same logarithmic discretization employed for the ML-MCTDH calculations~\cite{Wang_JCP_2001}. A total of 9, 12 and 64 bath modes were required to converge the uniform, symmetric, and asymmetric models, respectively. We used 16 and 64 beads for the simulations at 300 K and 50 K, respectively.

The classical rate constants $k_\text{cl}$ were calculated similarly to the RPMD rate constants by multiplying the classical TST rate $k^\ddag_\text{cl}$ with the transmission coefficient $\kappa_\text{cl}(t)$.%

The RPI simulations were carried out using the Nichols saddle-point search algorithm \cite{Nichols_JCPS_1990}; the number of replicas and the number of bath modes were converged to within graphical accuracy. Note that some of the authors have developed a version of the RPI method which includes the bath implicitly \cite{Litman_JCP_2022a}, but for consistency we used the same explicit (discretized) bath for the RPI calculations  as for the ML-MCTDH and RPMD calculations.

 To facilitate comparison with previous work,  we will sometimes present the "transmission coefficients" %
\begin{equation}
\begin{split}
    \overline\kappa(T) =\frac{ k(T)}{k^\ddag_\text{ha,cl}(T)},
    \end{split}\label{eq:kapaT_def}
\end{equation}
in place of the quantum, RPMD or classical rate constants $k(T)$, where  $k^\ddag_\text{ha,cl}$ is the harmonic approximation to the classical TST rate. 
To avoid confusion in what follows we always denote the coefficients of Eq.~\ref{eq:kapaT_def} with a bar, to distinguish them from the true transmission coefficients, $\kappa_\text{RPMD}$ and $\kappa_\text{cl}$, defined as in Eq.~\ref{eq:BC_rate}.

\section{Results\label{sec:results}}

\begin{figure}[h!]
\centering
\begin{subfigure}{.95\columnwidth}
  \centering
  \includegraphics[width=\columnwidth]{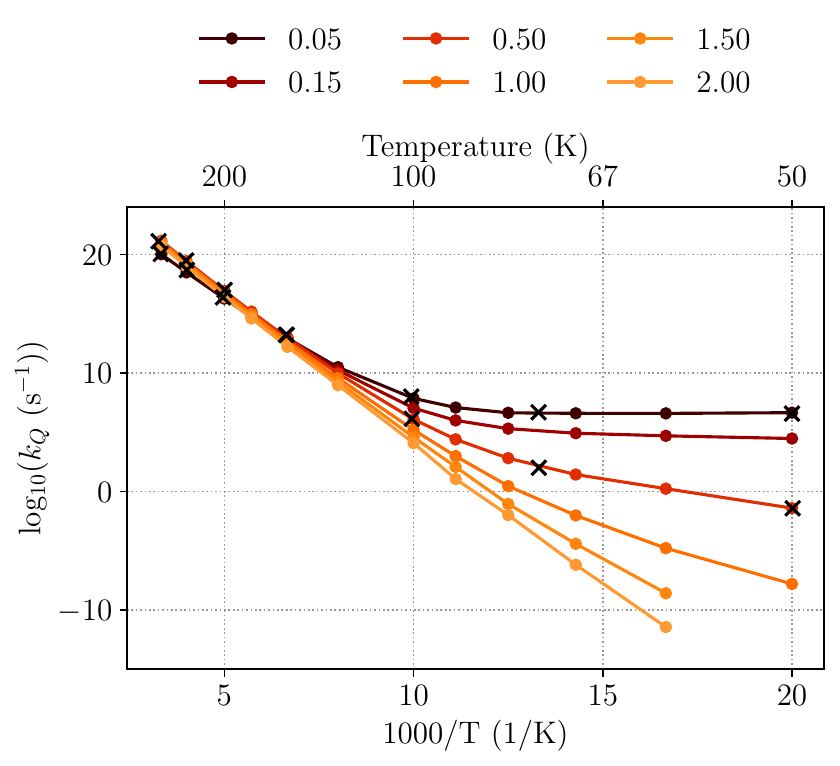}
  \subcaption{ Uniform friction}
\end{subfigure}

\begin{subfigure}{.95\columnwidth}
  \centering
  \includegraphics[width=\columnwidth]{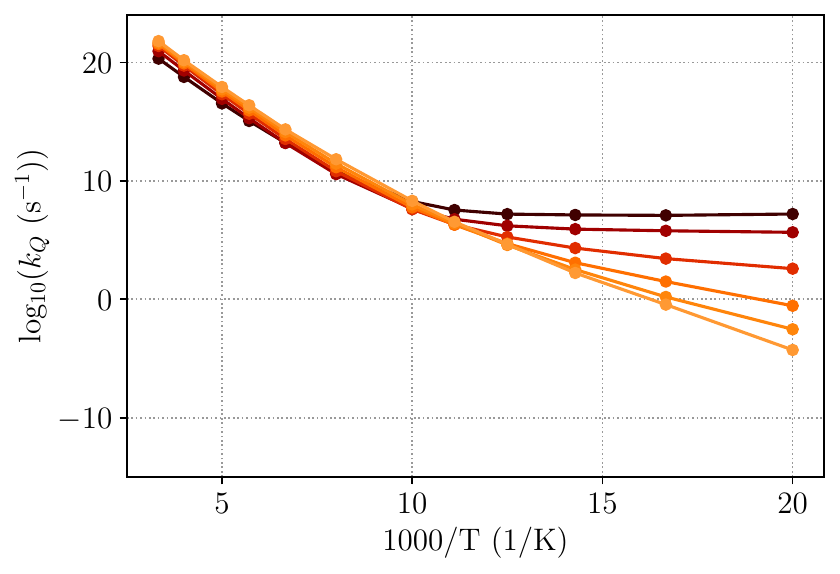}
  \subcaption{ Symmetric friction}
\end{subfigure}
\begin{subfigure}{.95\columnwidth}
  \centering
      \includegraphics[width=\columnwidth]{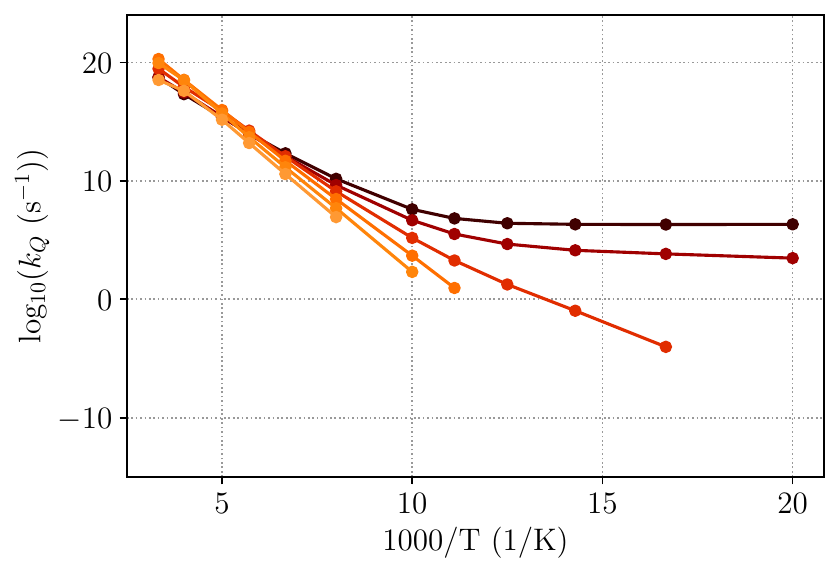}
  \subcaption{ Asymmetric friction }
\end{subfigure}
\caption{ML-MCTDH rate constants, $k_Q(T)$, for the three friction models of Tab.~I, for values of the reduced friction from $\eta_0/m \omega^\ddag$=0.05 (blue) to $\eta_0/m \omega^\ddag$=2.00 (pink).
Also shown are the QUAPI results of Ref.~\onlinecite{Topaler_JCP_1994} (black crosses).}
\label{fig:MCTDH_rates_arr}
\end{figure}

We start by presenting the ML-MCTDH quantum rate constants $k_\text{Q}(T)$ as a function of temperature.
Figure~\ref{fig:MCTDH_rates_arr} plots $k_\text{Q}(T)$ between 50~K and 300 K for $\eta_0/m \omega^\ddag$ between 0.05 and 2.00; note that the rate is plotted only when the flux-side correlation function shows a clear plateau.
The results for the uniform friction model are shown in panel a in Fig.~\ref{fig:MCTDH_rates_arr} along with the results reported by Topaler and Makri (TM) using the quasiadiabatic propagator path integral (QUAPI) \cite{Topaler_JCP_1994}. The two sets of results are in very close agreement.
A further comparison with reaction rates reported by 
Craig \cite{Craig_JCP_2007}  is presented in Fig.~S1 and shows equally good agreement.

The three models show an exponential decrease of the rate with temperature down to $\sim$ 100K where some of the simulations with the lowest friction reach a plateau characteristic of deep tunnelling. At low temperatures, the rates decrease monotonically with the increase of friction in all three models, showing that (as expected) dissipative effects inhibit quantum tunnelling, irrespective of the friction profile. At high temperatures, the rates follow a non-monotonic behaviour with varying friction, characteristic of Kramers-like behaviour \cite{Kramers_1940}.
In the following Sections, we analyze in more detail the dynamics at high and low temperatures. 

\begin{figure}
\centering
\begin{subfigure}{.95\columnwidth}
  \centering
  \includegraphics[width=\columnwidth]{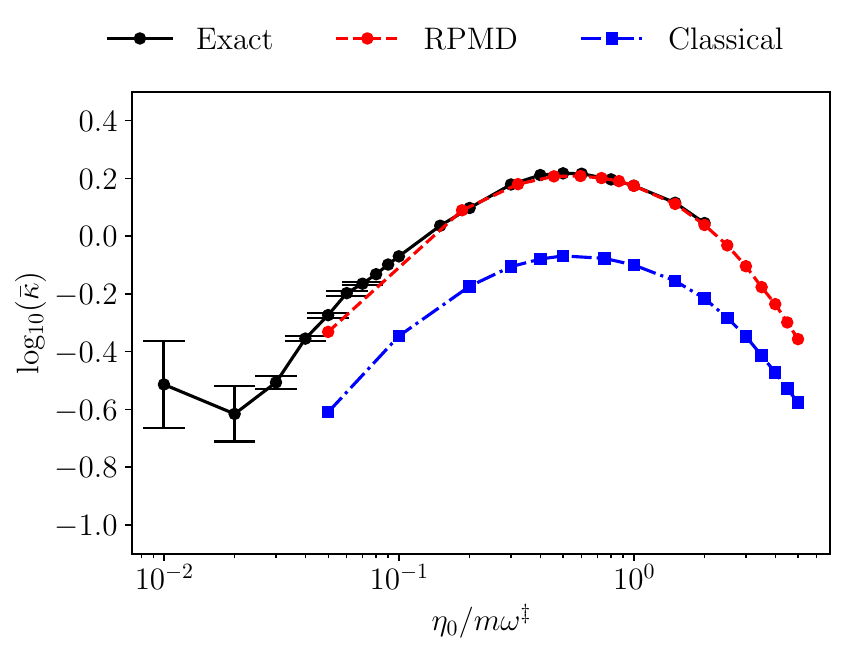}
  \caption{ Uniform friction }
\end{subfigure}
\begin{subfigure}{.95\columnwidth}
  \centering
  \includegraphics[width=\columnwidth]{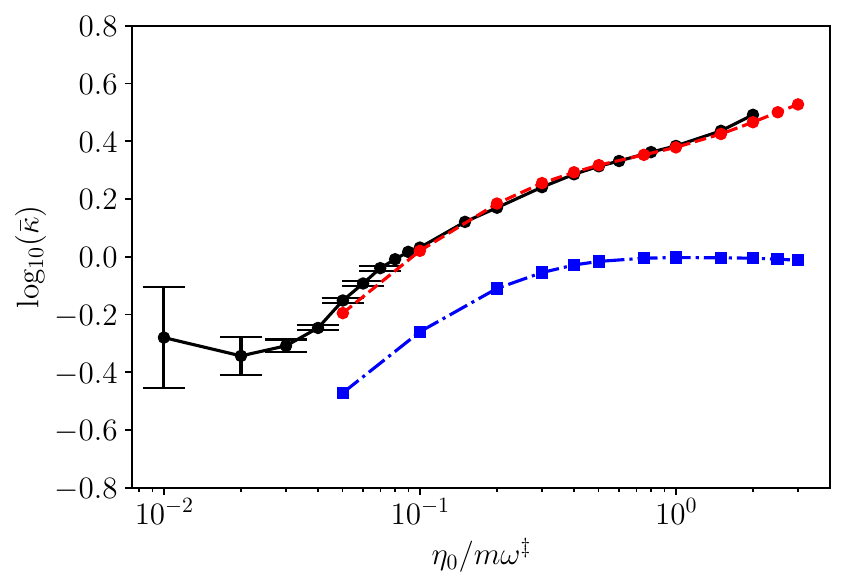}
  \caption{ Symmetric friction }
\end{subfigure}
\begin{subfigure}{.95\columnwidth}
  \centering
  \includegraphics[width=\columnwidth]{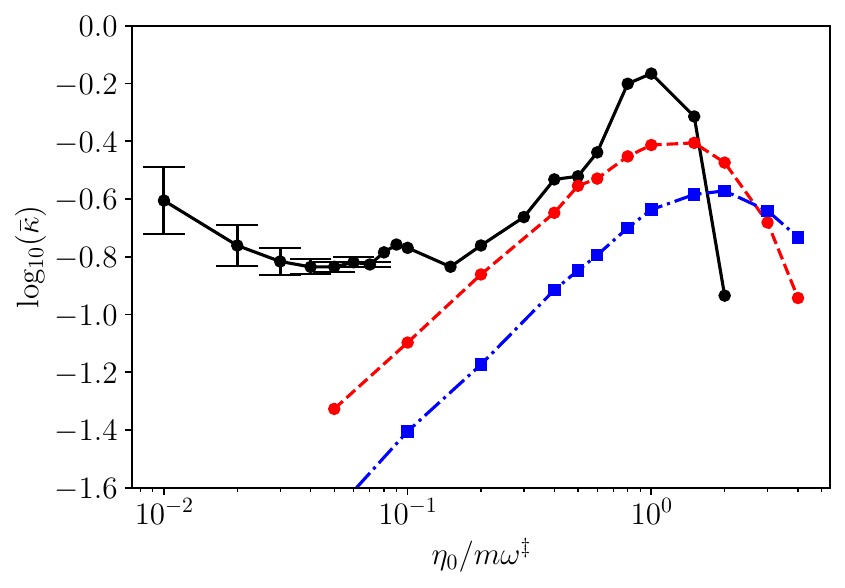}
  \caption{ Asymmetric friction }
\end{subfigure}
\caption{Transmission coefficients $\overline\kappa(T=300\,\text{K})$  for the \rev{(a) uniform friction, (b) symmetric friction, and (c) asymmetric friction models}, obtained from the corresponding ML-MCTDH,   RPMD 
and classical rate constants using Eq.~\ref{eq:kapaT_def}.
\rev{The error bars reported for the ML-MCTDH results with the lowest friction values correspond to the variation of the flux-side correlation functions during the last 100 fs of simulation (see Fig.~S2). Only error bars representing errors larger than 1\% are shown.}}
\label{fig:300K_results}
\end{figure}

\subsection{Quantum Effects at High Temperatures}
\subsubsection{Overview}

Figure~\ref{fig:300K_results}a shows the ML-MCTDH results obtained at 300~K for the uniform friction model.  These results are consistent with previous studies on this system, showing the characteristic Kramers' turnover \cite{Hanggi_RevModPhys_1990,Kramers_1940} separating the 
underdamped (energy-diffusion) and overdamped (spatial-diffusion) regimes. 
In the overdamped regime, $\eta_0/m\omega^\ddag >0.6 $,  the rate decreases monotonically with friction since the bath hinders the reactant from diffusing over the barrier. In the underdamped regime, 0.05 < $\eta_0/m\omega^\ddag$ < 0.3, the rate increases approximately linearly with the coupling strength since it is proportional to the rate at which energy can be transferred from the thermal fluctuations of the bath to the reaction coordinate.

\rev{  According to classical rate theories, at even weaker coupling, the rate should tend to zero. However, the ML-MCTDH simulations show another (purely quantum) regime, where the rate remains approximately constant within our uncertainties.
Representative flux-side time-correlation functions $c_\text{fs}(t)$ illustrating the dynamics in the three regimes are plotted in Fig.~S2, where the curves corresponding to the lowest friction couplings show the characteristic oscillation of coherent tunnelling.
 The transition from the tunnelling-dominated to the energy-diffusion regime at $\eta_0/m\omega^\ddag$ = 0.02  moves towards larger values of $\eta_0/m\omega^\ddag$ upon decreasing the temperature (see Fig.~S3), thus shrinking the energy-diffusion regime until it disappears at sufficiently low temperature.} \rev{Similar behaviour has been reported in recent studies by others \cite{Pollak_PRA_2024,Lindoy_NatCom_2023}. }
 \rev{The error bars reported in Fig.~\ref{fig:300K_results} could not be reduced further due to the appearance of bath recurrences (see Fig.~S4). However, we performed further ML-MCTDH calculations at 200 K, where this problem is less severe, and found that within the tunnelling-dominated regime, the reaction rates indeed increase when the friction is reduced (see Fig.~S5). Thus, in the low coupling limit, the increase of the dissipation strength reduces the rates by suppressing the (coherent) tunnelling \cite{Caldeira_AnPhys_1983, Grabert_PRL_1984, Weiss_book}. }

\begin{figure*}[t]
\centering
\begin{subfigure}{.328\textwidth}
  \centering
  \includegraphics[width=\linewidth]{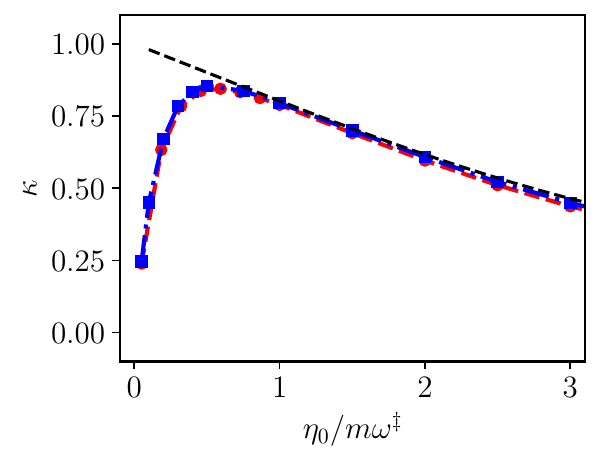}
  \caption{ Uniform friction }
\end{subfigure}
\begin{subfigure}{.328\textwidth}
  \centering
  \includegraphics[width=\linewidth]{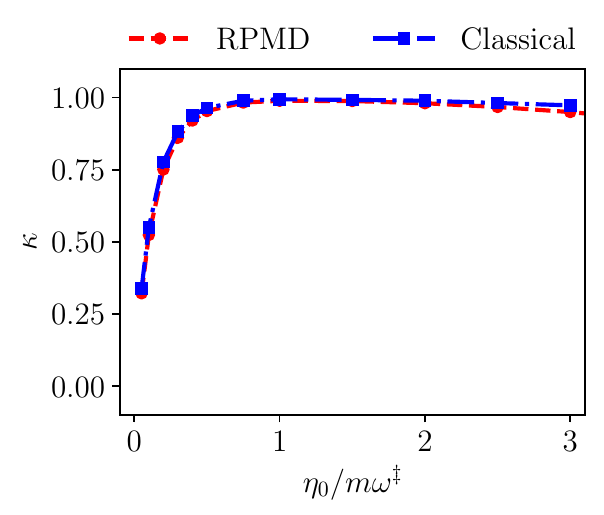}
  \caption{ Symmetric friction}
\end{subfigure}
\begin{subfigure}{.328\textwidth}
  \centering
  \includegraphics[width=\linewidth]{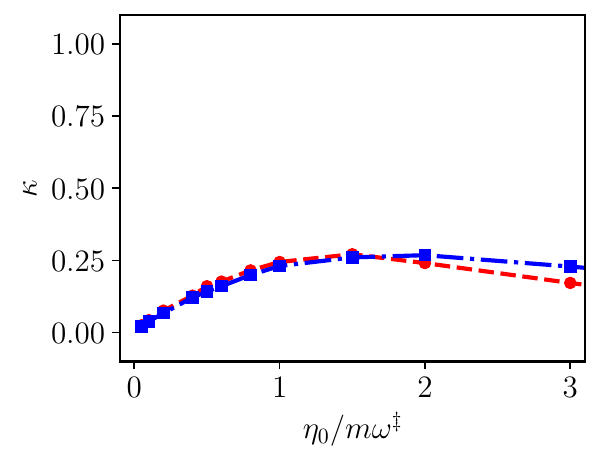}
  \caption{ Asymmetric friction }
\end{subfigure}
\caption{Transmission coefficients $\kappa(T=300\,\text{K})$, as defined in Eq.~\ref{eq:BC_rate}-\ref{eq:kp_RPMDTST_of_t}, calculated for the \rev{(a) uniform friction, (b) symmetric friction, and (c) asymmetric friction models}, using RPMD (red) and classical MD (blue). Panel (a) also shows the (classical) Grote-Hynes transmission factor\cite{Grote_JCP_1980, Grote_JCP_1981} (black dashed line).}
\label{fig:300K_recrossing}
\end{figure*}

Figure~\ref{fig:300K_results}b shows the ML-MCTDH results for the symmetric friction model. For $\eta_0/m\omega^\ddag>0.4$, the rate increases approximately linearly with $\eta_0/m\omega^\ddag$. This monotonic growth in the rate with $\eta_0/m\omega^\ddag$ arises because this model has zero friction in the region of the barrier (see Fig.~1); the addition of a small amount of friction to the barrier would cause an eventual turnover of the rate at sufficiently high $\eta_0/m\omega^\ddag$. Nonetheless, this monotonic growth depicted in Fig.~\ref{fig:300K_results}b cannot be explained away as a simple shift of the Kramers' turnover to higher friction values; we show below that it is a genuine quantum effect.

Figure~\ref{fig:300K_results}c shows the ML-MCTDH results for the asymmetric model. The ML-MCTDH calculations were much harder to converge for this model owing to the small system-bath coupling in the reactant well; tests suggest that, unlike the other two models, the ML-MCTDH calculations may not have fully converged with respect to the number of bath modes. \rev{Full convergence would require the development and optimization of a new ML-MCTDH tree structure which was not performed due to the prohibitive computational cost.} Despite these limitations, the ML-MCTDH calculations are sufficiently well converged to show that the asymmetric model gives rise to the same three kinetic regimes as the uniform model. The Kramers' turnover is shifted to a slightly higher value of $\eta_0/m\omega^\ddag~\sim$ 0.9, after which there is a much steeper drop in $\kappa(T)$ (which may be an artefact of the incomplete convergence just mentioned). The transition between the tunnelling-dominated and energy-diffusion regimes occurs at  
$\eta_0/m\omega^\ddag~\sim$ 0.1, a value five times larger than the one obtained for the uniform model. \rev{The similarity of the qualitative shapes of the uniform and asymmetric curves for the intermediate and large friction regimes suggests that the SDF in the asymmetric model could be approximated as an effective spatially independent friction, as in the effective Grote-Hynes theory of Voth \cite{Voth_JCP_1992}.}

To interpret the ML-MCTDH results, especially the monotonic growth of the rate with $\eta_0/m\omega^\ddag$ in the symmetric model, we compare with the results of classical and RPMD calculations. Figure~\ref{fig:300K_results} shows the RPMD rates calculated for the three models at 300K. 
For the uniform model, the RPMD simulations are in good agreement with the ML-MCTDH results in the underdamped and overdamped regimes. The RPMD calculations were difficult to converge in the tunnelling dominated regime ($\eta_0/m\omega^\ddag<0.05$), and in any case cannot be expected to work here, where the dynamics is dominated by real-time quantum coherence. The classical rates have the same Kramers' turnover as the RPMD rates, but are smaller by roughly a factor of two. The RPMD results for the symmetric model are also in good agreement with the ML-MCTDH results, showing the same monotonic increase in the rate with friction for $\eta_0/m\omega^\ddag>0.6$; the classical results, however, do not reproduce this feature, giving instead a plateau for $\eta_0/m\omega^\ddag>0.6$. For the asymmetric model, the RPMD and ML-MCTDH results differ appreciably. In the underdamped regime, these differences can be attributed to the neglect by RPMD of real-time coherence and coupling of the centroid to the Matsubara fluctuation modes \cite{Althorpe2021}; in the overdamped regime, while both methods show the same qualitative turnover behaviour, their difference is most likely due to the previously mentioned lack of convergence of the ML-MCTDH calculations.

\subsubsection{Analysis of Quantum Effects}
The qualitative differences between the RPMD and classical rates in Fig.~\ref{fig:300K_results} can be analysed in terms of the dynamical and statistical contributions to the rate defined by Eqs.~\ref{eq:BC_rate}--\ref{eq:kp_RPMDTST_of_t}, namely the transmission factors $\kappa(T)$ and the TST rates $k^\ddag(T)$. We can further factorise the latter into
\begin{equation}
\label{eq:kp_TST_2}
{k}_\text{RPMD}^\ddag (T) = \bigg(\frac{k_B T}{2\pi m} \bigg)^{1/2} e^{-\beta \Delta{\cal A}_\text{RPMD}^\ddag}
\frac{\langle \delta(q_{0}^{c}-\overline q_0)\rangle}{\langle h(q_0^\ddag-q_{0}^{c})\rangle},
\end{equation}
where
\begin{equation}
\begin{split}
\Delta{\cal A}_\text{RPMD}^\ddag  = {\cal A}_\text{RPMD}(q_0^\ddag)-{\cal A}_\text{RPMD}(\overline q_0),
 \end{split}
\end{equation}
and
\begin{equation}
\begin{split}
{\cal A}_\text{RPMD}(q_0)  = -{1\over \beta} \ln{\langle \delta(q_{0}^{c}-q_0)\rangle}
\end{split}
\end{equation}
is the free energy difference needed to move the centroid from the bottom of the well, $\overline q_0$, to the barrier top, $q_0^\ddag$.

\begin{figure}%
    \centering
    \begin{subfigure}{\columnwidth}
  \centering
  \includegraphics[width=.95\linewidth]{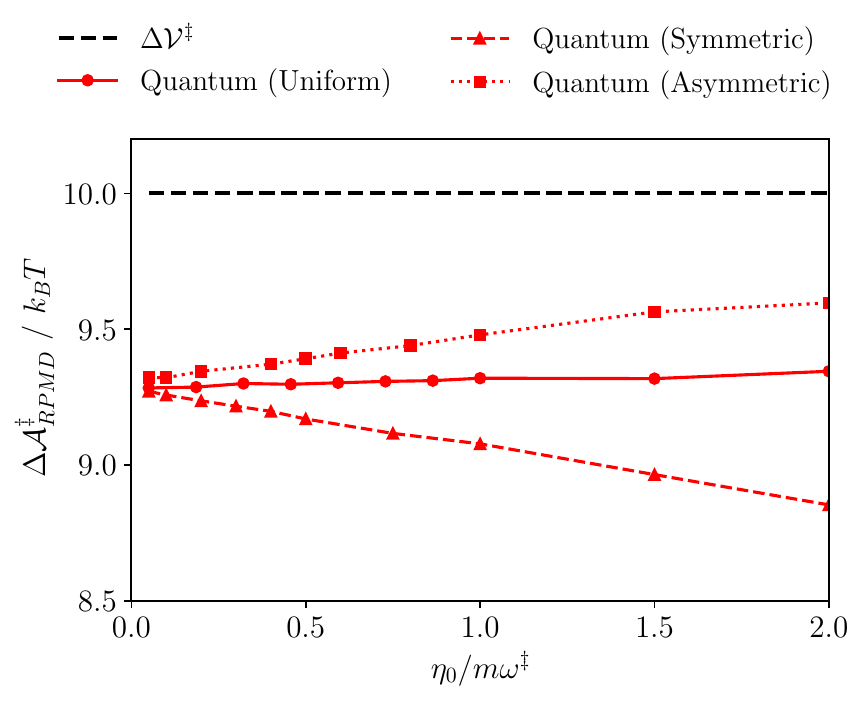}
     \caption{  Free energy at 300K.}
\end{subfigure}
    \begin{subfigure}{1.0\columnwidth}
  \centering
  \includegraphics[width=.41\linewidth]{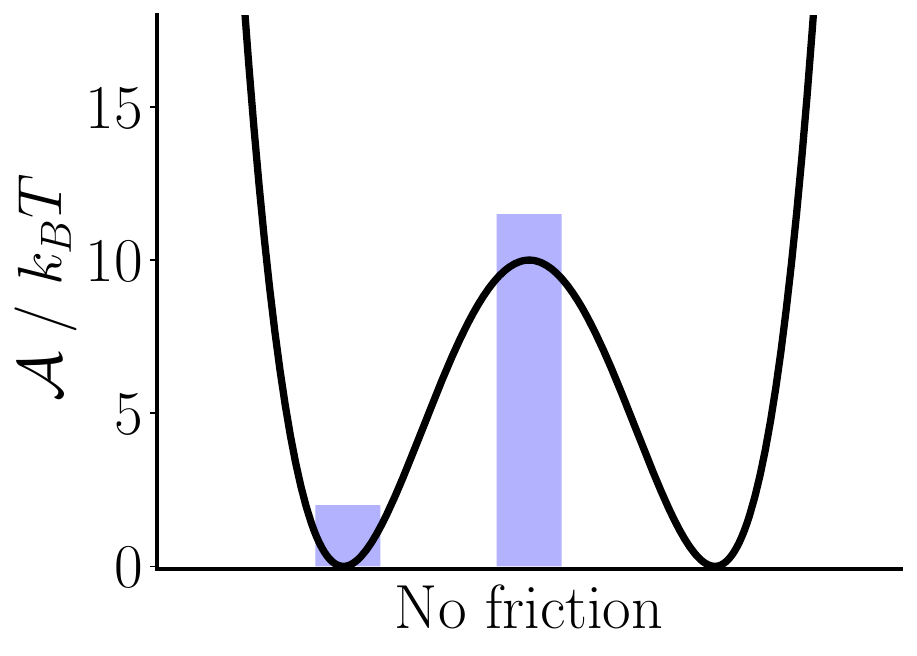}
  \includegraphics[width=.41\linewidth]{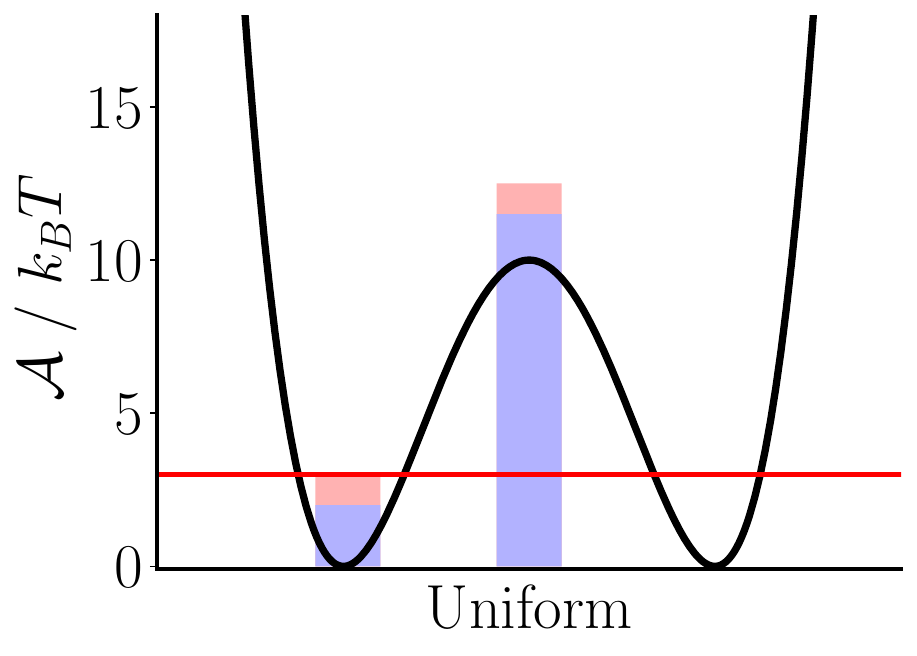}
  \includegraphics[width=.41\linewidth]{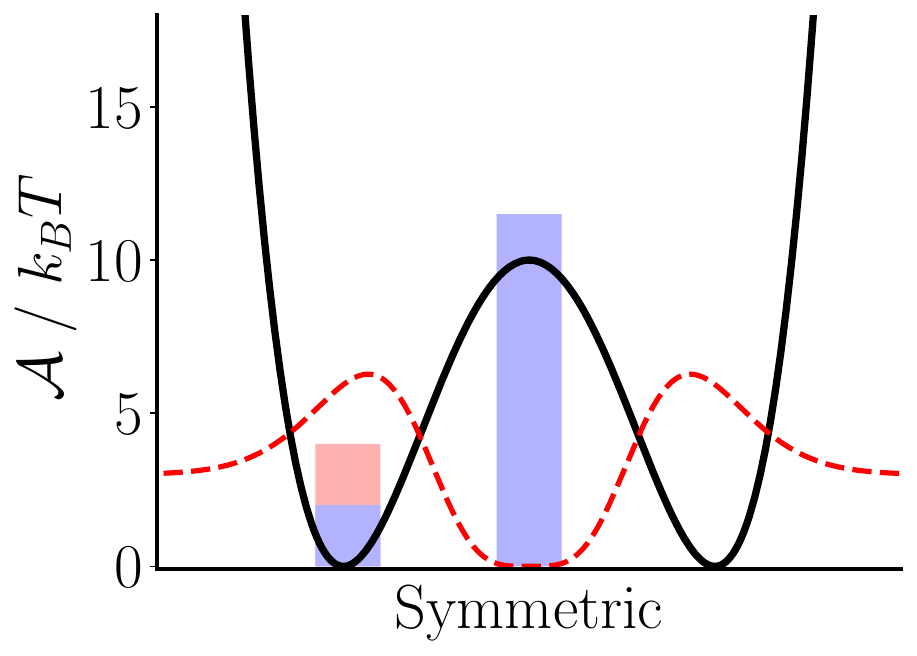}
  \includegraphics[width=.41\linewidth]{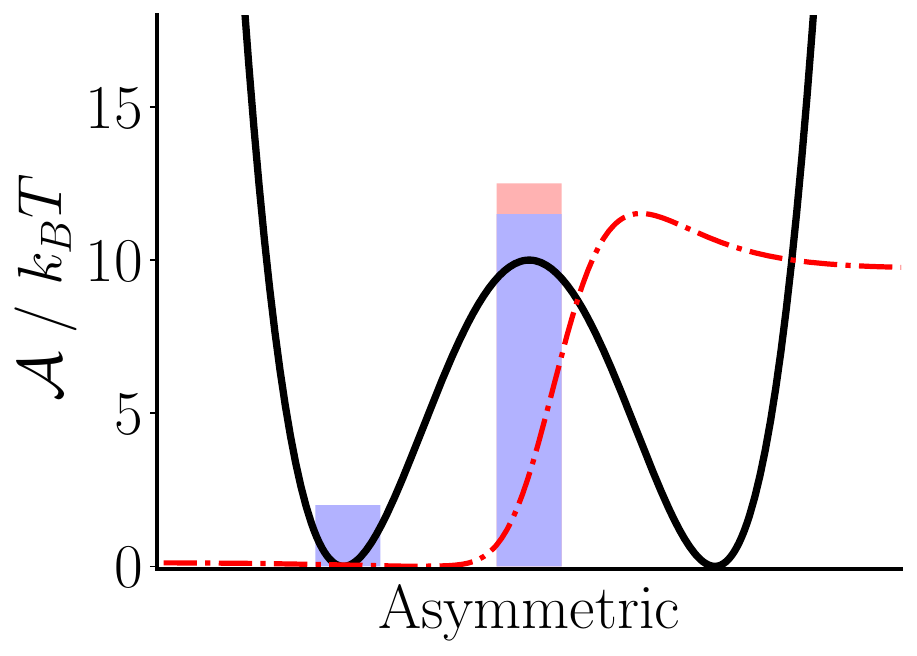}
  \caption{ Free energy contributions.}
\end{subfigure}

  \caption{Analysis of the dependence of the quantum free-energy on the bath coupling strength $\eta_0/m\omega^\ddag$ at 300 K, showing (a)  the reaction free-energy $\Delta{\cal A}_\text{RPMD}^\ddag$ for the three friction models, and (b) the free-energy at the barrier and in the well, decomposed into contributions from thermal quantum fluctuations of the pure system (blue) and the system-bath coupling (pink). $\Delta{\cal V}^\ddag = V(q^\ddag)-V(\overline q_0)$.   Friction contributions are estimated using Eq.~\ref{eq:harmonic_free_energy} and multiplied by a factor of four to ease visualization.}
  \label{fig:fe_300K}
\end{figure}

The transmission coefficients $\kappa(T)$ are plotted in Figure~\ref{fig:300K_recrossing}, and the free energy differences $\Delta{\cal A}_\text{RPMD}^\ddag$ in Figure~\ref{fig:fe_300K}.
Fig.~\ref{fig:300K_recrossing} shows that $\kappa_\text{RPMD}(T)$ is very close to $\kappa_\text{cl}(T)$ for the uniform and symmetric models. In other words, there are almost no quantum effects in the recrossing dynamics for these two models over the entire range $\eta_0/m\omega^\ddag>0.05$ for which RPMD reproduces the   
ML-MCTDH rates (in Fig.~\ref{fig:300K_results}). The small differences between $\kappa_\text{RPMD}(T)$ and $\kappa_\text{cl}(T)$ for the asymmetric model (Fig.~\ref{fig:300K_recrossing}c) are perhaps a sign that delocalisation between the zero friction and high friction halves of the potential affects the recrossing dynamics.%

The strong quantum effects that cause the rates for the symmetric model to increase monotonically with $\eta_0/m\omega^\ddag$ for $\eta_0/m\omega^\ddag>0.5$ (see Fig.~\ref{fig:300K_results}b)  are thus caused by the dependence of the quantum free-energy gap $\Delta{\cal A}_\text{RPMD}^\ddag$ on $\eta_0/m\omega^\ddag$. It is well known \cite{Weiss_book,Litman_JCP_2022a} that the quantum free energy of a system-bath model depends strongly on $\eta_0/m\omega^\ddag$: the system ring-polymer modes orthogonal to the centroid couple to their counterparts in the bath, which increases the effective polymer-spring force constants, causing the ring polymers to shrink as though the fluctuations were occurring at an effectively higher temperature, or as if the system's mass were increased. Within the harmonic approximation, and for 
a bath with Ohmic spectral density,
the free energy of the uniform model at the bottom of the well is \cite{Weiss_book,Litman_JCP_2022a}$^{,}$
\footnote{We assume here that $P$ is odd.}
\begin{equation}
\label{eq:harmonic_free_energy}
\begin{split}
{\cal A}_\text{RPMD}(\overline q_0) = \frac{1}{\beta} \ln \left[ \prod ^{(P-1)/2}_{l=0}   \omega_l^2 + \frac{|\omega_l| \eta_0}{m} + (\omega_\text{well})^2\right] \\ + V(q_0^\ddag) + \text{consts},
\end{split}\end{equation}
\nid where $\omega_l= 2\omega_P \sin(|l|\pi/P)$ are the free ring polymer normal mode frequencies and $\omega_\text{well}$ is the frequency at the 
bottom of the well. %
The classical limit of Eq.~\ref{eq:harmonic_free_energy} is obtained by taking $P=1$, and since $\omega_{l=0}=0$, the classical free energy is indeed friction and mass independent.  Eq.~\ref{eq:harmonic_free_energy} clearly shows that ${\cal A}$ increases monotonically with the increase of friction. This dependence, which also applies to anharmonic potentials, can be rationalized in two equivalent ways: either as a renormalization of the system mass or as an increase of the quantum fluctuations.

In Fig.~\ref{fig:fe_300K}a,  the classical and quantum reaction free energies at 300 K are presented.
$\Delta{\cal A}_\text{RPMD}^\ddag$ at  $\eta_0$=0 is smaller than the classical one due to the zero-point energy at the reactant well.
For finite values of friction, $\Delta{\cal A}_\text{RPMD}^\ddag$ stays approximately constant for the uniform model, whereas it decreases (increases) for the symmetric  (asymmetric) model. The different behaviour between the uniform and non-uniform friction models is a direct consequence of the increase of quantum free energy with friction (as approximately described by Eq.~\ref{eq:harmonic_free_energy} using a local harmonic approximation). While the free energy at the reactant basin and transition state region increase comparably in the model with uniform friction, due to the shape of the friction profile in the symmetric model, the free energy at the reactant well increases more rapidly than at the transition state for this case. This difference leads to the monotonic decrease of the free energy barrier with friction shown in panel (a) and schematically represented in panel (b). \rev{Our results are consistent with those previously reported in Ref. \onlinecite{Navrotskaya_ChemPhys_2006} based on QUAPI and centroid-molecular-dynamics \cite{CMD2} simulations on similarly symmetric SDF profiles}. The asymmetric model presents vanishing friction at the reactant basin and finite friction at the top of the barrier. An analogous argument explains the increase of $\Delta{\cal A}_\text{RPMD}^\ddag$ with the increase of friction observed in this case (see curve in Fig.~\ref{fig:fe_300K}a and sketch in Fig.~\ref{fig:fe_300K}b).

\subsection{Quantum Effects at Low Temperatures}

Another consequence of the dependence of the free energy on the system-bath coupling  is that the instanton cross-over temperature is reduced below the value of $T_c^\circ$, given in Eq.~\ref{eq:Tc},  to \cite{Grote_JCP_1980,Pollak_JCP_1986,Litman_JCP_2022a}
  \begin{equation}
      \label{eq:Tc_renorm}
 T_c(\eta_0)= T_c^\circ \left[\left(\frac{\tilde{\eta}(\omega_0)}{2m \omega^\ddag}\right)^2+1\right]^\frac{1}{2}-\frac{\tilde{\eta}(\omega_0)}{2m \omega^\ddag},
\end{equation}
\nid where $\omega_0$ is the free-energy barrier frequency and the friction is evaluated at the top of the barrier\footnote{Note that this equation has to
be solved self-consistently.}. For the DW1 potential,  the cross-over temperature reduces from 115 K for the uncoupled system to 92 K for $\eta_0/m\omega^\ddag=1$. The ML-MCTDH results display a smooth transition between the high-temperature and low-temperature regimes (see Fig.~S3). To simplify the discussion we focus on the results at 50 K, where the system remains in the deep tunnelling regime across the entire range of friction.

\begin{figure}
\centering
\begin{subfigure}{\columnwidth}
  \centering
  \includegraphics[width=.95\linewidth]{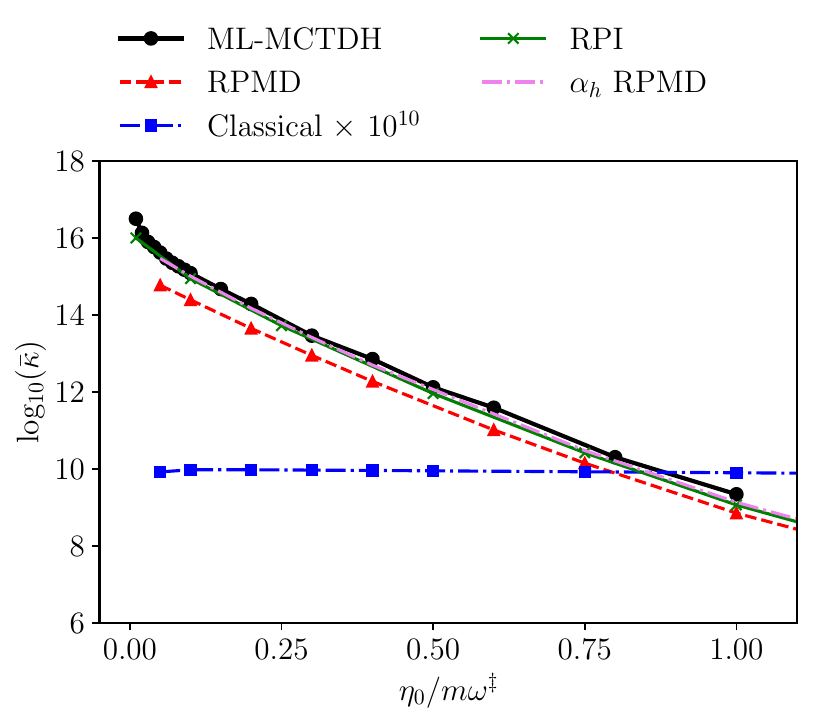}
  \caption{ Uniform friction  }
\end{subfigure}
\begin{subfigure}{\columnwidth}
  \centering
  \includegraphics[width=.95\linewidth]{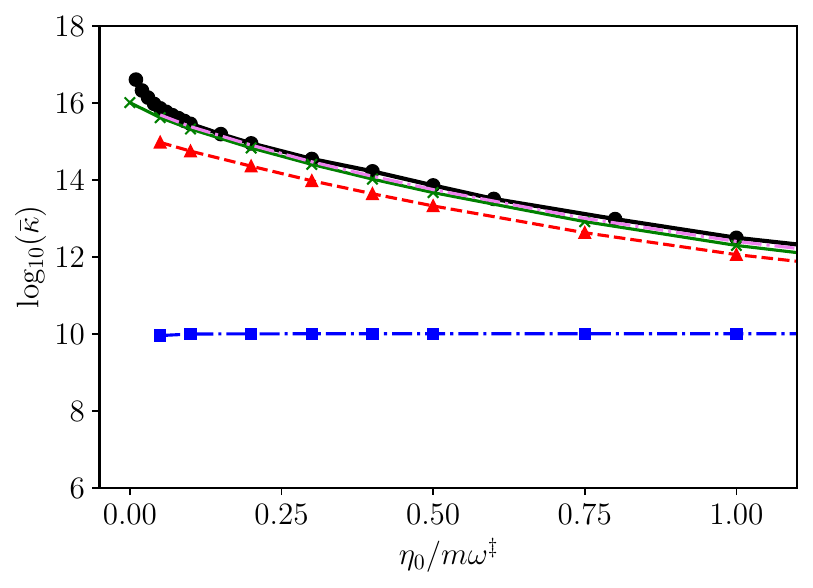}
  \caption{ Symmetric friction }
  \label{fig:sub3}
\end{subfigure}
\begin{subfigure}{\columnwidth}
  \centering
  \includegraphics[width=.95\linewidth]{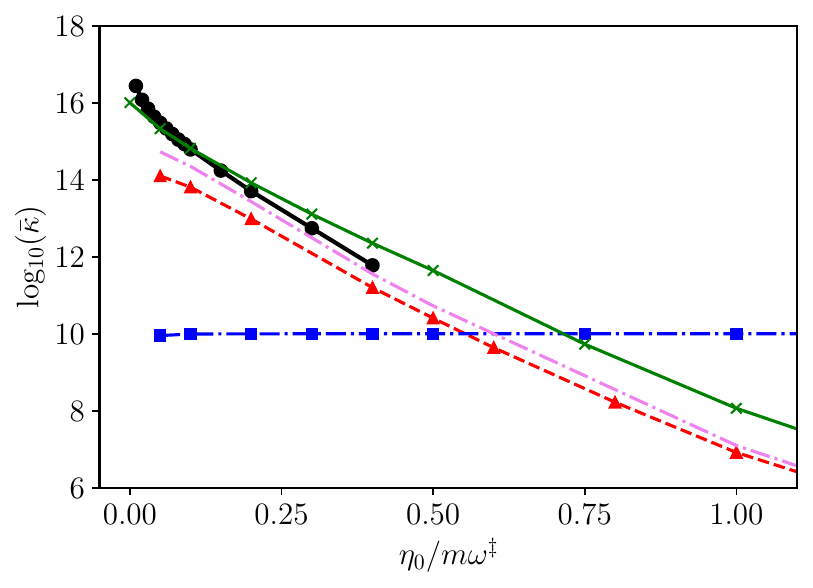}
  \caption{ Asymmetric friction}
\end{subfigure}

\caption{Transmission coefficients $\overline\kappa(T=50\,\text{K})$  for the \rev{(a) uniform friction, (b) symmetric friction, and (c) asymmetric friction models}, obtained from the corresponding ML-MCTDH,   RPMD, and RPI rate constants using Eq.~\ref{eq:kapaT_def}. The `$\alpha_h$ RPMD' coefficients were obtained by multiplying the RPMD rate constants by the $\alpha_h(\beta)$ correction factor of Eq.~\ref{alphah}.}
\label{fig:050K_results}
\end{figure}

Figure~\ref{fig:050K_results} compares the reaction rates at 50 K using ML-MCTDH, RPMD, RPI and classical MD. 
In all models,  the reaction rate decreases with friction in a qualitatively similar way, in contrast with the high-temperature results. The Kramers turnover is not present since it is exponentially suppressed with temperature \cite{Hanggi_ZPB_1987,Topaler_JCP_1994, PGH}.
The decrease of the rate with friction in the symmetric and asymmetric models is less pronounced than for the uniform coupling, suggesting
that even in the deep tunnelling regime, quantum dynamics are particularly sensitive to how the frictional forces change in the vicinity of the barrier top. This difference is also observed in the
centroid-free energies along the reaction pathway (see  Fig.~S6 in the supporting information).

Since the reaction is dominated by deep tunnelling at this temperature, the classical rates are orders of magnitude smaller than the quantum rates. 
RPI results are \rev{within a factor of two} of the ML-MCTDH ones proving that in the deep-tunnelling regime, real-time quantum dynamics plays a minor role, except at vanishing friction strength. \rev{In this limit, namely for $\eta_0/m\omega^\ddag<0.05$, the ML-MCTDH rates show a marked increase with the decrease of friction, in agreement with simulations reported using hierarchical equations of motion (HEOM) for an effective two-state spin-boson model \cite{Shi_JCP_2011}.}
\rev{The larger discrepancy observed for the asymmetric model, reaching a factor of four at the largest computed value of friction, is probably due to the lack of convergence of the ML-MCTDH results, as discussed in the high-temperature results. }
RPMD  systematically underestimates the RPI reaction rates,  which is typical for a symmetric reaction barrier, and can be traced back to the different treatment of the unstable mode, $\lambda_0$, in RPI and the harmonic TST version of RPMD
\cite{Richardson_JCP_2009}. The ratio between the RPI and harmonic RPMD-TST rates can be shown to be\cite{Richardson_JCP_2009}
\begin{equation}
\label{alphah}
\alpha_h(\beta) = \frac{2\pi}{\beta \hbar |\lambda_0|},
\end{equation}
which for the uniform model varies from 4.1 at $\eta_0/m\omega^\ddag=0.1$ to 1.9 at $\eta_0/m\omega^\ddag=1.0$.

\section{ Conclusions}\label{sec:Conclusions}

We have investigated the effects on Kramers'-type reactions of making the friction spatially-dependent. We carried out accurate (ML-MCTDH) and approximate (RPMD, RPI and classical) simulations on two very different spatially-dependent friction profiles (called "symmetric" and "asymmetric"). We find that the spatial dependence introduces strong quantum effects into the rate constants for both models, over the full range of friction strengths considered (which, for a uniform friction strength would encompass the full range of the Kramers' curve).  However, only at very low overall friction strengths can these effects be attributed to real-time quantum dynamics. At higher friction strengths, the quantum effects are found to be static, reflecting the changes in the quantum free-energy profiles that result when the friction is made spatially-dependent. We also find tentative evidence (from the tests on the asymmetric profile) that quantum dynamical effects might play a role in systems with a \rev{steep variation} between low and high friction, but these results are inconclusive (since we do not know whether the ML-MCTDH calculations have converged to the exact quantum rates for this friction profile).

Methodologically, these results imply that spatially-dependent friction models behave similarly to the commonly used system-bath models with spatially-independent friction. Thus, at low overall friction, accurate quantum methods are needed to capture the real-time coherence. Here, we used the ML-MCTDH method; hopefully, these results will be useful to others \rev{ for benchmarking the variety of approximate methods, e.g. linearized semiclassical theory \cite{Liu_JCP_2009,Liao_JCP_2002}, that have been successfully applied to the uniform friction case}. \rev{As mentioned in the Introduction, QUAPI has already been applied to position-dependent friction baths by Navrotskaya and Geva \cite{Navrotskaya_ChemPhys_2006}, and
we expect that the HEOM method \cite{HEOM_I,Chen_JCP_2009}
could be similarly extended}. However, once the friction is large enough to remove most of the real-time coherence (which is still in the energy-diffusion-limited Kramers' regime) then RPMD is, by construction, able to capture all the quantum free-energy effects and thus will usually work well in this regime (except perhaps for friction profiles with {steep variations}---see above). This is useful to know, as RPMD is typically orders of magnitude cheaper than accurate quantum methods. 

There are a variety of systems which could be modelled realistically by spatially-dependent friction and which could thus be treated using the approaches used in this article. In the low friction regime, we mention in particular optical cavities \cite{Fiechter_JPCL_2023, Lindoy_NatCom_2023, Mandal_ChemRev_2023} where real-time quantum effects are thought to be important; at higher frictions, a version of RPMD has already been used to study the diffusion of light particles in metallic environments \cite{Dou_2023}. 

\phantom{lorem ipsum dolor sit amet lorem ipsum dolor sit amet lorem ipsum dolor sit amet lorem ipsum dolor sit amet lorem ipsum dolor sit amet lorem ipsum dolor sit amet}

\section*{Supplementary Material}

See the supplementary material for additional figures including ML-MCTDH rate constants at intermediate temperatures, ML-MCTDH flux-side correlation functions, and centroid free energy profiles along the reaction pathways.

\appendix

\begin{acknowledgments}

Y.L. acknowledges funding from the Deutsche Forschungsgemeinschaft (DFG, German Research Foundation), project number 467724959. P.L. and M.R. thank George Trenins for his valuable discussions and for providing an alternative (more efficient) code for system-bath RPMD. R.M. gratefully acknowledges the INDACO platform, which is a project of High Performance Computing at the Università degli Studi di Milano, for the computational resources allocated at the CINECA HPC center.
\end{acknowledgments}

\section*{Data Availability}

The data that support the findings of this study are openly
available at \url{https://gitlab.com/litman90/quantum_spatially_varying_friction}.

\bibliographystyle{aipnum4-1}

\end{document}


\preprint{AIP/123-QED}

\title{Supporting Information: Quantum rates in dissipative systems with spatially varying friction}

\author{Oliver Bridge \orcidlink{0009-0008-6835-5082}}%
\affiliation{Yusuf Hamied Department of Chemistry,  University of Cambridge,  Lensfield Road,  Cambridge,  CB2 1EW, UK}

\author{Paolo Lazzaroni}%
\affiliation{MPI for the Structure and Dynamics of Matter, Luruper Chaussee 149, 22761 Hamburg, Germany \looseness=-1}

\author{Rocco Martinazzo \orcidlink{0000-0002-1077-251X}}%
\affiliation{Department of Chemistry, Università degli Studi di Milano, Via Golgi 19, 20133 Milano, Italy}

\author{Mariana Rossi \orcidlink{0000-0002-3552-0677}}%
\affiliation{MPI for the Structure and Dynamics of Matter, Luruper Chaussee 149, 22761 Hamburg, Germany \looseness=-1}

\author{Stuart C. Althorpe}%
\affiliation{Yusuf Hamied Department of Chemistry,  University of Cambridge,  Lensfield Road,  Cambridge,  CB2 1EW, UK}

\author{Yair Litman \orcidlink{0000-0002-6890-4052}}%
\email[Email: ]{yl899@cam.ac.uk}
\affiliation{Yusuf Hamied Department of Chemistry,  University of Cambridge,  Lensfield Road,  Cambridge,  CB2 1EW, UK}

\maketitle

\section{Additional Figures}

\begin{figure}[h!]
    \centering
    \includegraphics[width=0.5\columnwidth]{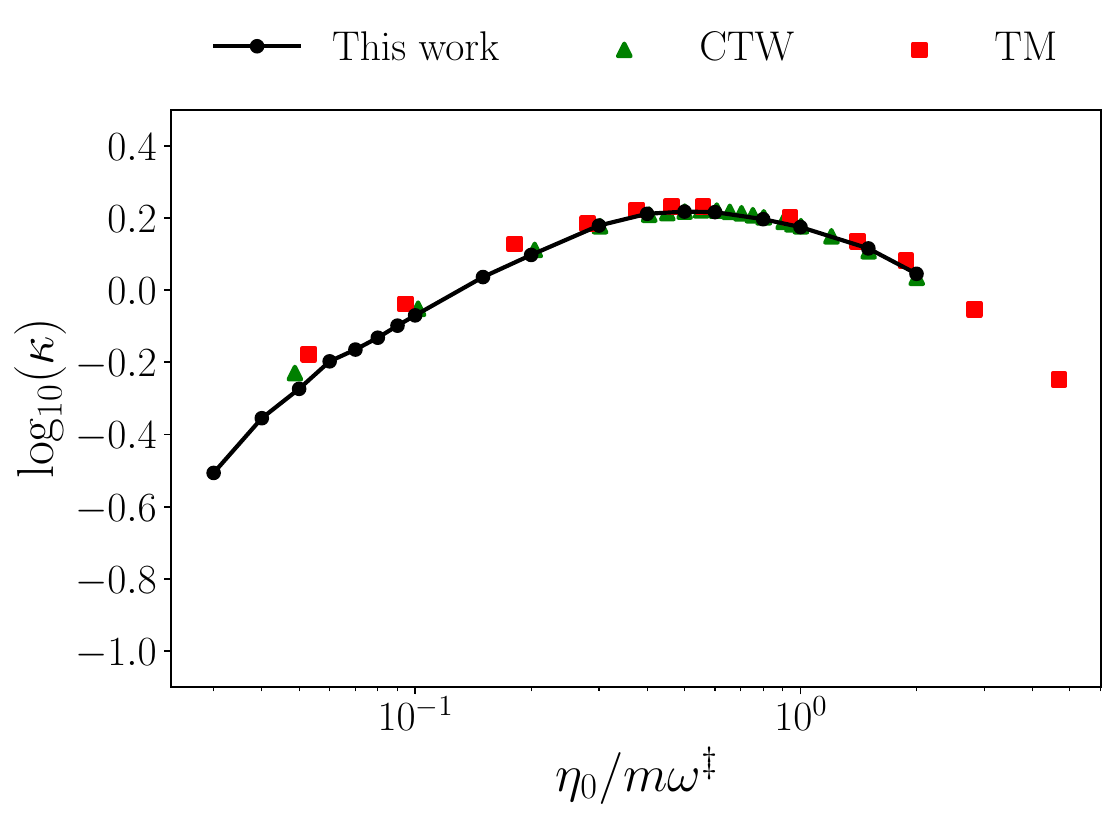}
    \caption{Comparison of the transmission factors at 300K for the uniform friction model
obtained in this work, by Craig, Wang and Thoss (CTW)\cite{Craig_JCP_2007}, and by Topaler and Makri (TM)\cite{Topaler_JCP_1994}.}
    \label{fig:Craig_comparison}
\end{figure}

In  Fig.~\ref{fig:Craig_comparison}, we compare the transmission factor reported in this work for the uniform friction model at 300K with the results previously reported in the literature by Craig, Wang and Thoss (CTW) \cite{Craig_JCP_2007}, and Topaler and Makri (TM) \cite{Topaler_JCP_1994}.
The three datasets are in perfect agreement. 

\begin{figure}[h!]
\centering
\begin{subfigure}{.32\textwidth}
  \centering
  \includegraphics[width=\linewidth]{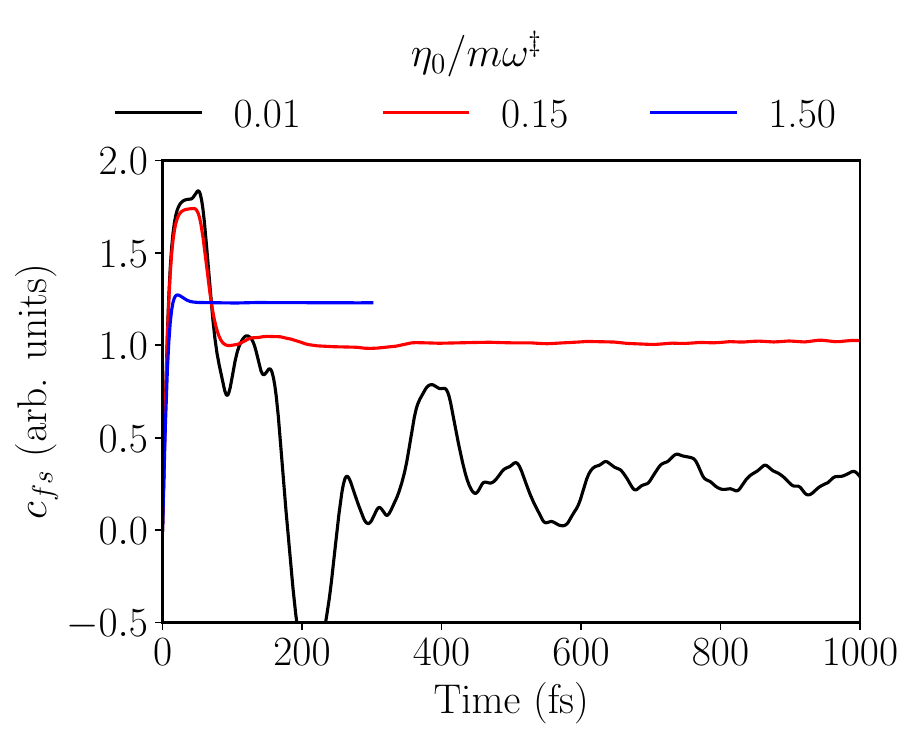}
  \caption{ Uniform friction }
\end{subfigure}
\begin{subfigure}{.32\textwidth}
  \centering
  \includegraphics[width=\linewidth]{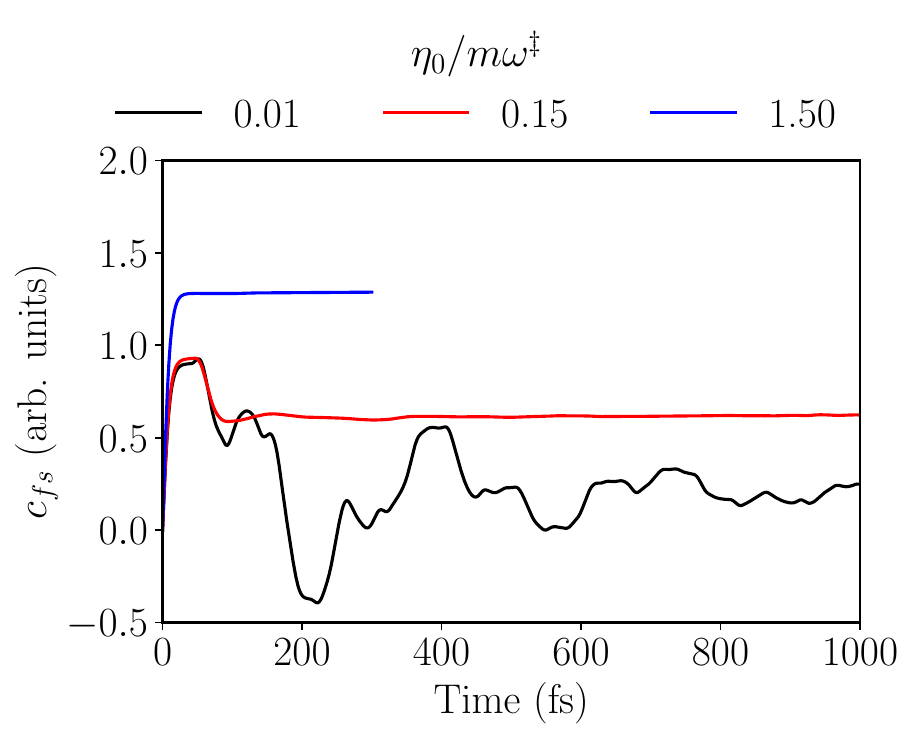}
  \caption{ Symmetric friction}
\end{subfigure}
\begin{subfigure}{.32\textwidth}
  \centering
  \includegraphics[width=\linewidth]{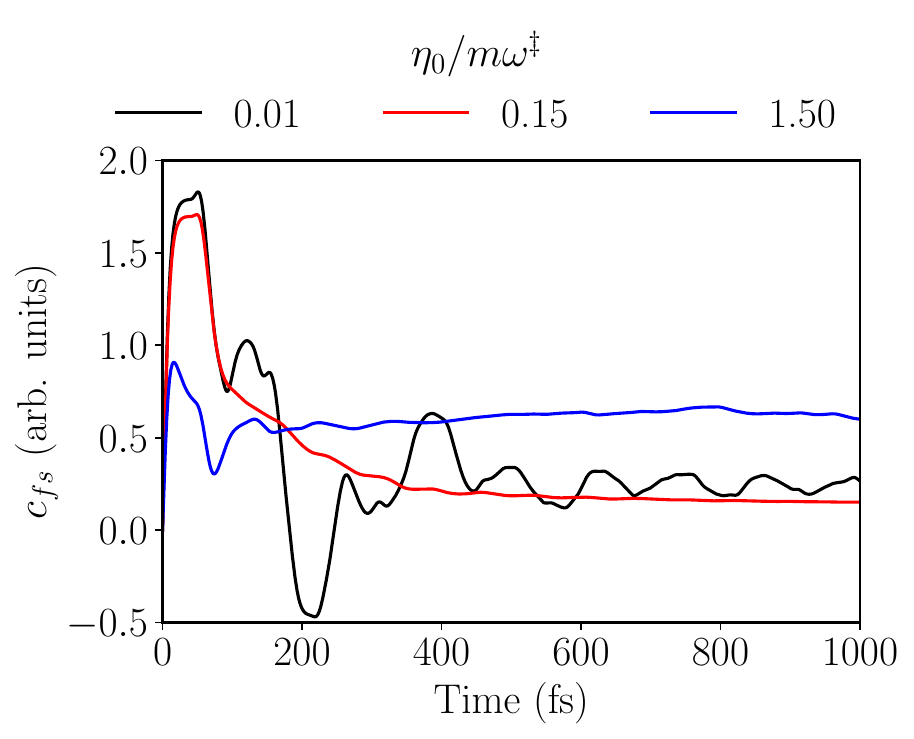}
  \caption{ Asymmetric friction}
\end{subfigure}
\caption{Flux-side correlation function, $c_\text{fs}(t)$, obtained with ML-MCTDH at 300K for selected friction values.}
\label{fig:Cfs_300K}
\end{figure}

In Fig. \ref{fig:Cfs_300K}, we show the flux-side correlation function obtained with multi-layer multi-configuration time-dependent Hartree (ML-MCTDH) at representative friction values at 300K. In all cases,  the coherent tunnelling-dominated regime at low friction can be identified by the presence of oscillation that extends over hundreds of femtoseconds. 

\begin{figure}[h!]
\centering
\begin{subfigure}{.32\textwidth}
  \centering
  \includegraphics[width=\linewidth]{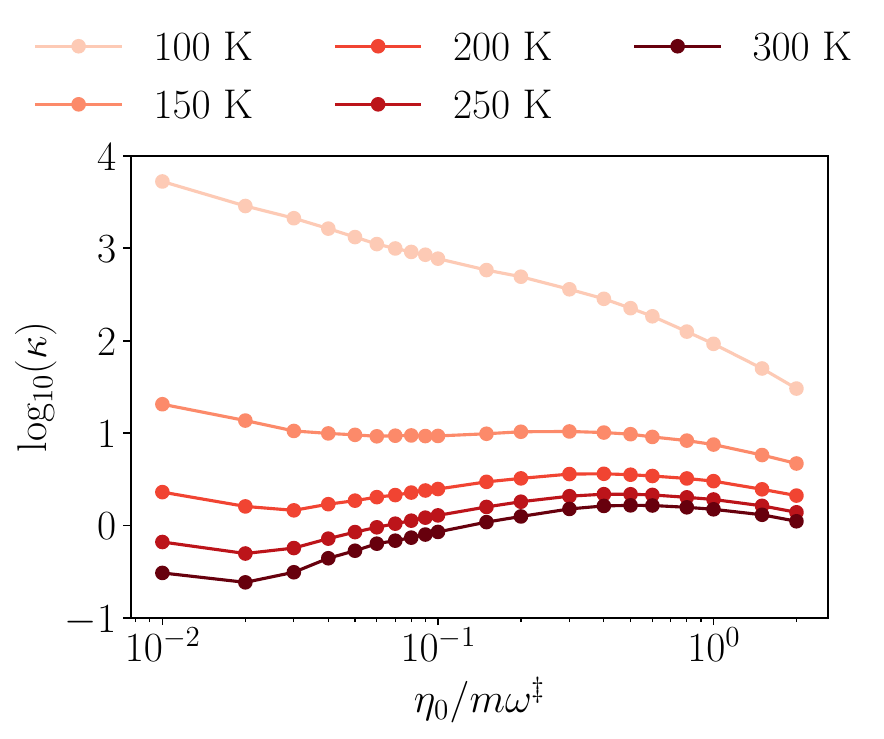}
  \caption{ Uniform friction}
\end{subfigure}
\begin{subfigure}{.32\textwidth}
  \centering
  \includegraphics[width=\linewidth]{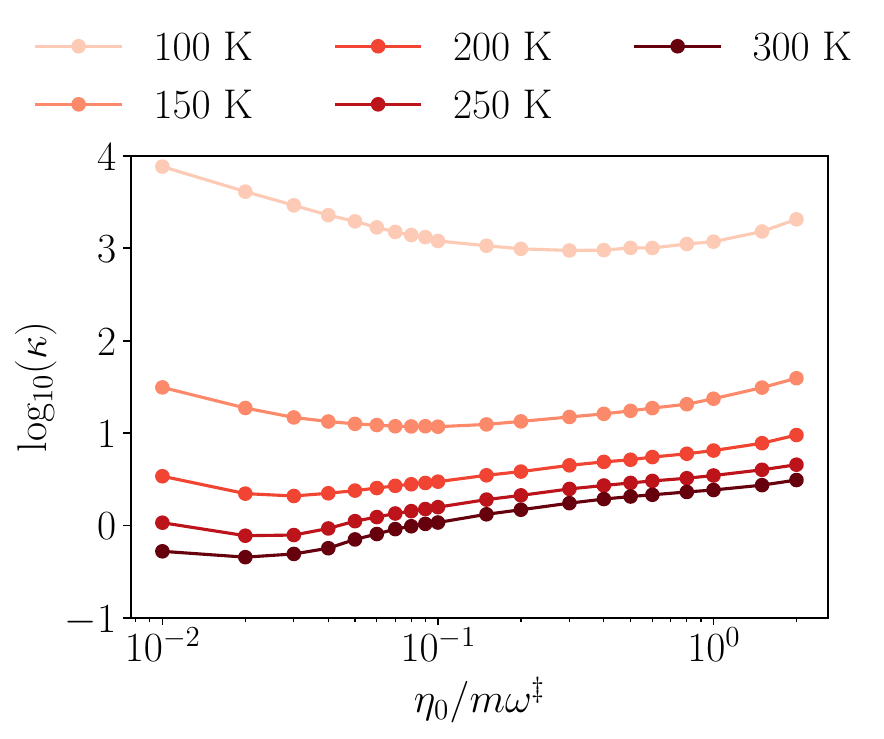}
  \caption{ Symmetric friction}
\end{subfigure}
\begin{subfigure}{.32\textwidth}
  \centering
  \includegraphics[width=\linewidth]{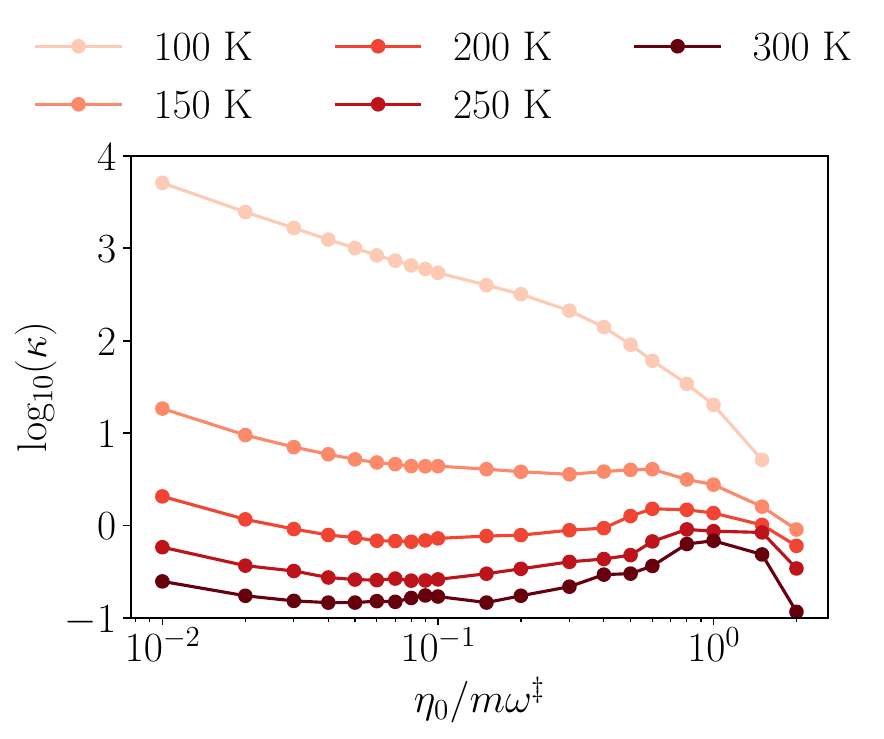}
  \caption{ Asymmetric friction}
\end{subfigure}

\caption{Transmission factors (Eq. 24 in the main text) obtained from the ML-MCTDH simulations.%
}
\label{fig:Kramers_results}
\end{figure}

In  Fig. \ref{fig:Kramers_results}, we report the transmission factors obtained with 
ML-MCTDH simulations across a wide range of temperatures.

\begin{figure}[h!]
    \centering
    \includegraphics[width=0.5\columnwidth]{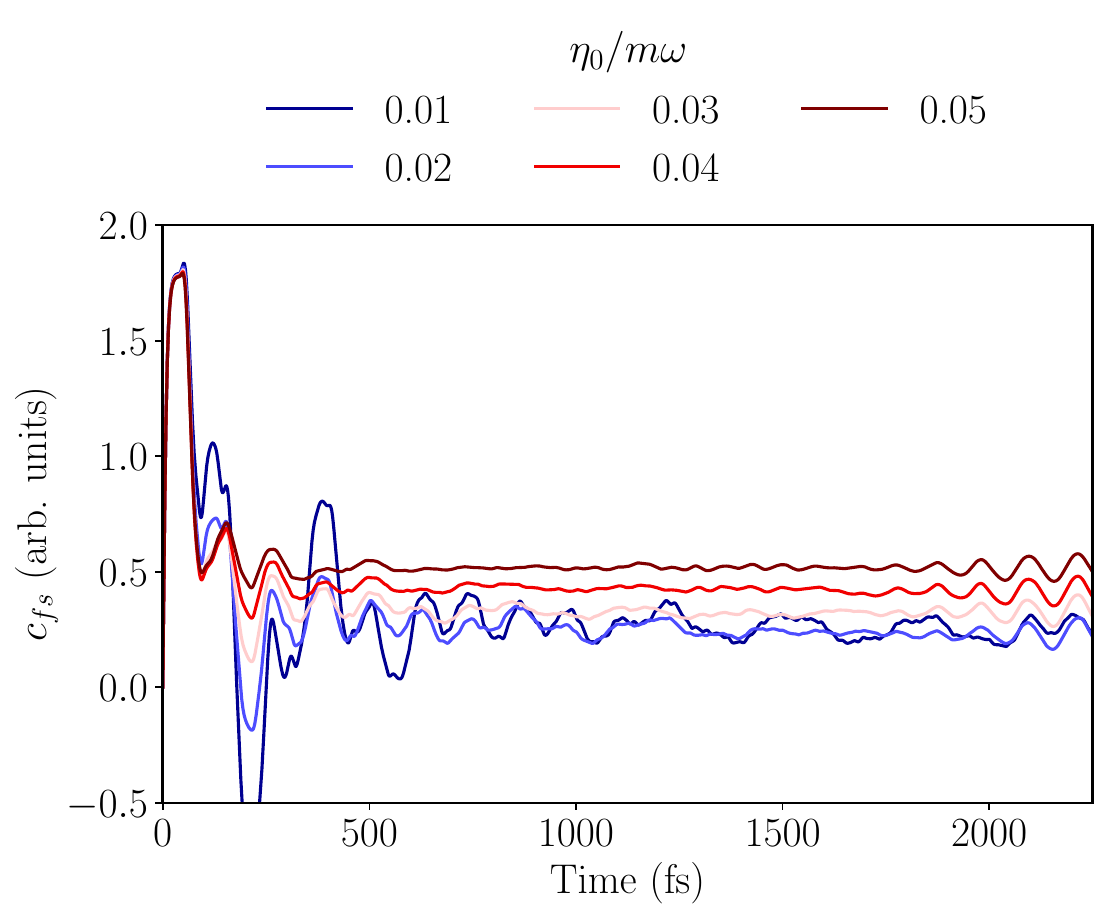}
    \caption{ Extended flux-side correlation function, $c_\text{fs}(t)$, obtained with ML-MCTDH at 300K for the uniform model. The oscillations that build close to 2000 fs are attributed to bath recurrences. }
    \label{fig:300K_cfs_long}
\end{figure}

Fig. \ref{fig:300K_cfs_long} shows the flux-side correlation function, $c_\text{fs}(t)$, for the uniform model at 300 K over an extended period of time (up to 2250 fs), where the artefacts induced by the bath recurrences can be observed. To increase the bath recurrence time,
 the development and optimization of a new ML-MCTDH tree structure with a larger number of bath modes would be required. 

\begin{figure}[h!]
\centering
\begin{subfigure}{.32\textwidth}
  \centering
  \includegraphics[width=\linewidth]{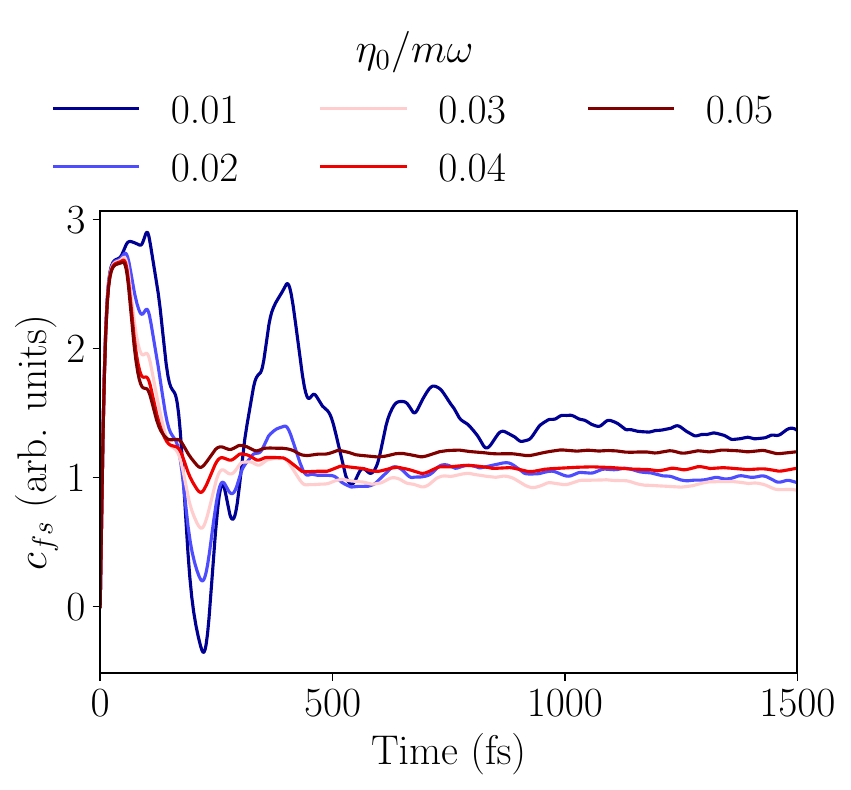}
  \caption{ Uniform friction }
\end{subfigure}
\begin{subfigure}{.32\textwidth}
  \centering
  \includegraphics[width=\linewidth]{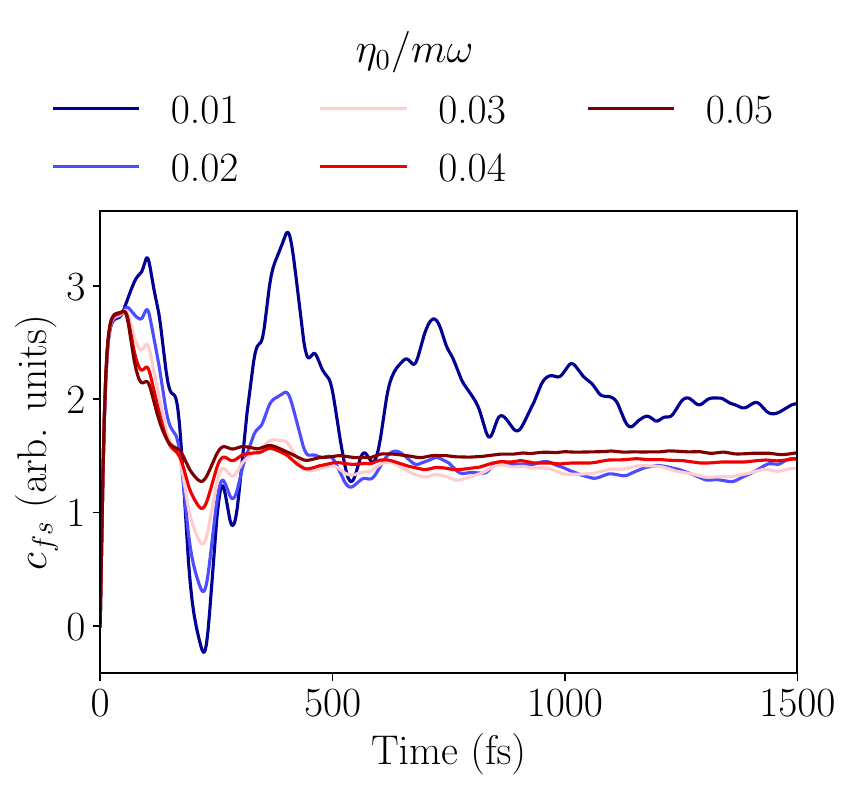}
  \caption{ Symmetric friction}
\end{subfigure}
\begin{subfigure}{.32\textwidth}
  \centering
  \includegraphics[width=\linewidth]{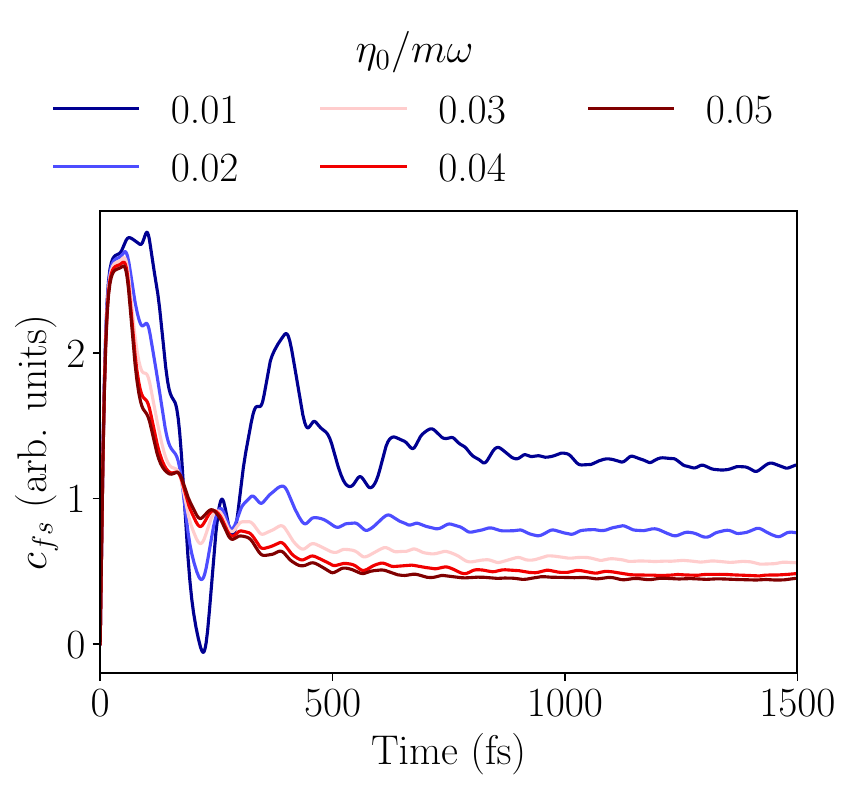}
  \caption{ Asymmetric friction}
\end{subfigure}
\caption{Flux-side correlation function, $c_\text{fs}(t)$, obtained with ML-MCTDH at 200K for selected friction values. In all cases, the largest long-time limit value is obtained for the lowest values of friction}
\label{fig:Cfs_200K}
\end{figure}

In Fig. \ref{fig:Cfs_200K}, we report the flux-side correlation functions calculated at 200 K for the three models considered. It can be observed that the long-time limit value of $c_\text{fs}(t)$, which is proportional to the rate constant in this narrow range of friction values, reaches its maximum for the lowest values of friction.

\begin{figure}[H]
    \centering
    \begin{subfigure}{0.32\columnwidth}
        \centering
         \includegraphics[width=\linewidth]{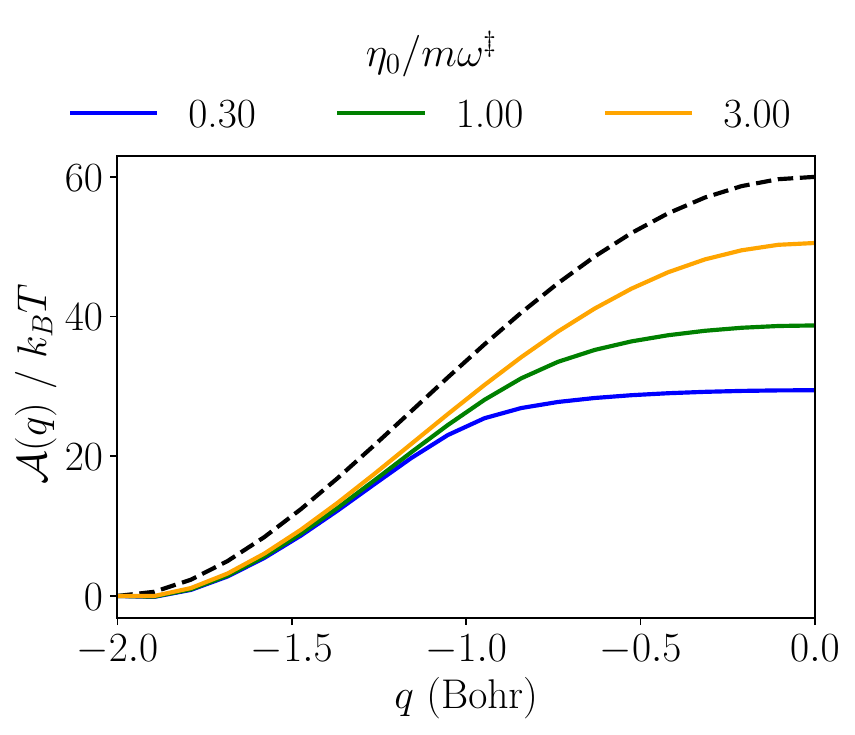}
         \caption{ Uniform friction}
     \end{subfigure}
     \begin{subfigure}{0.32\columnwidth}
         \centering
          \includegraphics[width=\linewidth]{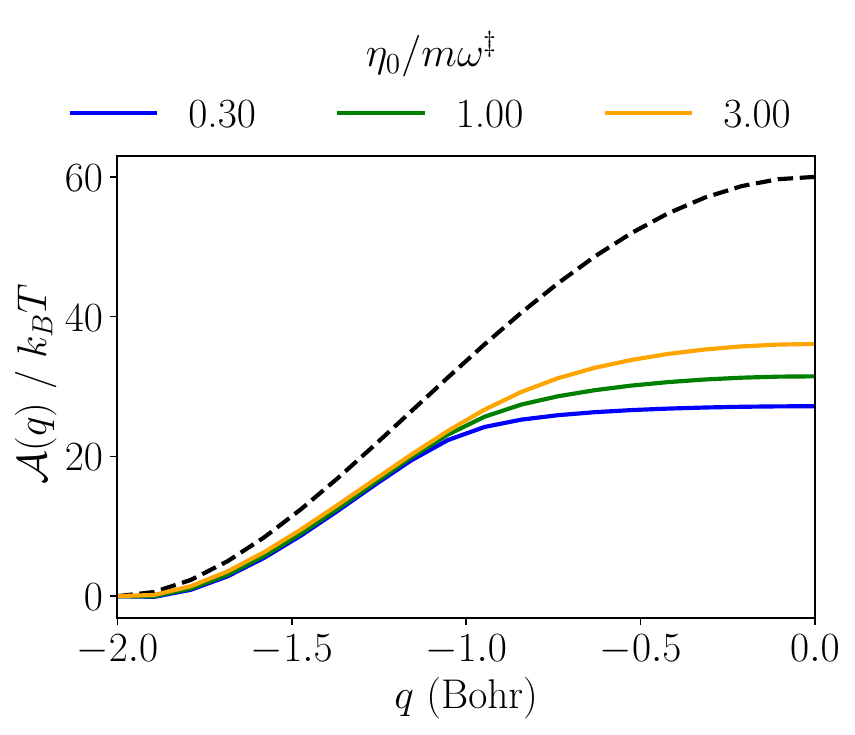}
          \caption{ Symmetric friction}
     \end{subfigure}
     \begin{subfigure}{0.32\columnwidth}
          \centering
          \includegraphics[width=\linewidth]{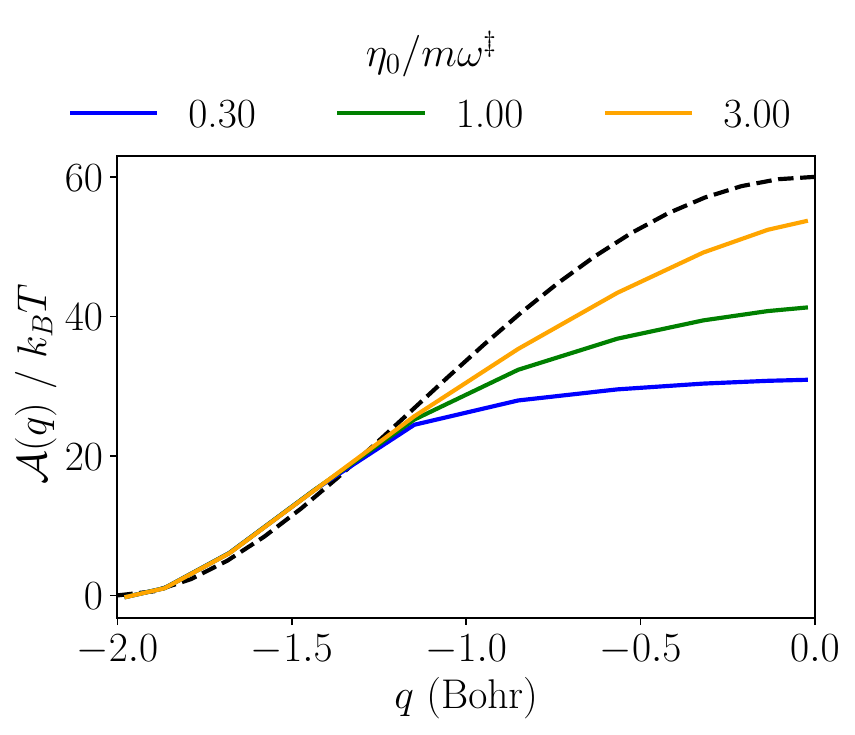}
          \caption{ Asymmetric friction}
     \end{subfigure}
     
         \caption{Centroid free energy along the reaction pathway at 50K for $\eta_0/m\omega_b$=0.3 (blue),
         $\eta_0/m\omega_b$=1.0 (green), and
         $\eta_0/m\omega_b$=3.0 (orange).  The classical free energy is depicted by a black dashed line.       
         \label{fig:centroid_fe_050K} }
\end{figure}

In  Fig. \ref{fig:centroid_fe_050K}, we present centroid-free energies along the reaction pathway at 50 K. In the calculations of the uniform model, the free energy in the vicinity of the barrier top flattens out for the lowest friction values, an indication of the ring-polymer delocalization \cite{Drechsel-Grau_Angew_2014,Cendagorta_PCCP_2016},
whereas it remains fairly parabolic for the largest values of friction. This qualitative change can be traced back to the appearance of instanton-like geometries at lower friction strengths. Since the cross-over temperatures are 71 K and 53 K for 
$\eta_0/m\omega^\ddag=0.3$ and $\eta_0/m\omega^\ddag=3.0$, respectively, at 50 K the system is only in the deep-tunneling regime for the former case. In the symmetric model calculations, the changes of the free energy profiles with friction are milder since the instanton geometry expands across areas of relatively small friction. Similar to what is observed at high temperatures,  the results of the asymmetric model resemble the uniform case.

